\documentclass[10pt,prb,aps,twocolumn,showpacs,
superscriptaddress]{revtex4}
\usepackage{graphicx}
\usepackage{dcolumn}% Align table columns on decimal point
\usepackage{bm}% bold math
\usepackage{amssymb}
\newcommand{\dn}{\downarrow}
\newcommand{\up}{\uparrow}
\newcommand{\ua}{\uparrow}
\newcommand{\da}{\downarrow}

\begin{document}

\title{Stochastic series expansion algorithm \\ for the $S=1/2$ XY model with
four-site ring exchange}

\author{Roger G. Melko}
\affiliation{Department of Physics, University of California,
Santa Barbara, California 93106}

\author{Anders W. Sandvik}
\affiliation{Department of Physics, Boston University,
590 Commonwealth Ave., Boston, Massachusetts 02215.}
\affiliation{Department of Physics, University of California,
Santa Barbara, California 93106}

\date{\today}

\begin{abstract}
We describe a stochastic series expansion (SSE) quantum Monte Carlo method
for a two-dimensional $S=1/2$ XY-model (or, equivalently, hard-core bosons 
at half-filling) which in addition to the standard pair interaction $J$ 
includes a four-particle term $K$ that flips spins on a square plaquette. The 
model has three ordered ground state phases; for $K/J \alt 8$ it has 
long-range $xy$ spin order (superfluid bosons), for $K/J \agt 15$ it
has staggered spin order in the $z$ direction (charge-density-wave), and 
between these phases it is in a state with columnar order in the bond and 
plaquette energy densities. We discuss an implementation of directed-loop 
updates for the SSE simulations of this model and also introduce a 
``multi-branch'' cluster update which significantly reduces the autocorrelation
times for large $K/J$. In addition to the pure J-K model, which in the 
$z$ basis has only off-diagonal terms, we also discuss modifications of 
the algorithm needed when various diagonal interactions are included.
\end{abstract}


\maketitle

\section{Introduction}

In the ongoing quest to explore possible ground states and quantum phase
transitions in quantum condensed matter systems 
(fermions, bosons, or quantum spins), 
numerical studies are important for establishing the true nature of the phases
and transitions of relevant model Hamiltonians. 
In particular, recent interest in ``exotic'' phenomena has focused attention
on models with frustrated or competing interactions, in which interplay 
between adjacent ordered phases often gives rise to interesting effects.
\cite{senthil,exoticrefs,RSrk,balents} For classical models, Monte Carlo 
simulations in combination with finite-size scaling can be used very 
successfully in studies of a wide range of systems with and without 
frustration. However, only a limited class of quantum models
are amenable to such studies, as the infamous sign problem prohibits
large-scale quantum Monte Carlo (QMC) studies of frustrated 
antiferromagnetic spin systems and fermions in more than one 
dimension. It is therefore important to search for 
non-sign-problematic quantum models, possibly with competing interactions,
that display complex ground state phase diagrams and can be efficiently
studied using Monte Carlo simulations.
Although not all possible types of ground states and quantum phase
transitions may be realizable within this class of Hamiltonians, it is likely
that many insights into the low-temperature physics of quantum matter can 
still be gained in this way. 
Constructing optimized and efficient quantum Monte Carlo 
algorithms for such candidate Hamiltonians is hence an important task.

\begin{figure}
\includegraphics[width=8cm]{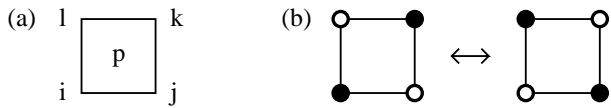}
\caption{(a) Labeling convention for the indices of an operator $O_{ijkl}$ 
acting on the corners of a plaquette. The label $p$ refers to the whole 
plaquette, so that $O_p \equiv O_{ijkl}$. (b) The two plaquette configurations
between which the $K$-term can act; open and solid circles correspond to up 
and down spins, respectively.}
\label{fig:labels}
\end{figure}

In this paper, we present the details of a stochastic series expansion (SSE)
algorithm that we have developed for large-scale QMC studies of a 
two-dimensional (2D) $S=1/2$ XY model with an added four-site ring-exchange term
(the method can be easily generalized for three-dimensional 
systems\cite{roger3d}). 
Defining the following bond and plaquette operators;
\begin{eqnarray}
B_{ij} & = & S^+_iS^-_j + S^-_iS^+_j = 2(S^x_iS^x_j + S^y_iS^y_j), 
\label{bond} \\
P_{ijkl} & = & S^+_iS^-_jS^+_kS^-_l + S^-_iS^+_jS^-_kS^+_l, 
\label{plaquette}
\end{eqnarray}
the J-K Hamiltonian is given by
\begin{equation}
H = -J\sum\limits_{\langle ij\rangle} B_{ij}
    -K\sum\limits_{\langle ijkl\rangle} P_{ijkl},
\label{ham}
\end{equation}
where $\langle ij\rangle$ denotes a pair of nearest-neighbor sites on a 2D
square lattice and $\langle ijkl\rangle$ are sites on the corners of a 
plaquette, as illustrated in Fig.~\ref{fig:labels}(a). The plaquette-flip 
$P_{ijkl}$ is only a subset of all the possible cyclic exchanges among 
four spins and corresponds to retaining only the purely $x$- and $y$-terms;
it has a non-vanishing matrix element only between the two spin states with 
alternating (staggered) spins on the corners of the plaquette,
as illustrated in Fig.~\ref{fig:labels}(b). 
In the standard way, the J-K Hamiltonian (\ref{ham}) 
can also be considered as a half-filled hard-core boson model, where up and 
down spins correspond to filled and empty sites and $J$ is the 
nearest-neighbor hopping. We will frequently use terminology referring to 
this boson representation. With the negative sign in front of the 
plaquette-term ($K>0$), 
the J-K model can be studied using QMC methods without a sign 
problem (the sign of the $J$-term is actually irrelevant in this regard). 
In this model, there is no frustration in the conventional sense, i.e., 
antiferromagnetic interactions on lattice loops with an odd number of 
links (which leads to sign problems). However, the $J$- and $K$-terms 
individually favor different types of ground states, which leads to 
interesting competition effects at intermediate $K/J$. 

The J-K model was recently found to exhibit three different ordered ground 
states as a function of the ratio $K/J$ of the four-site ($K$) and two-site ($J$) 
terms.\cite{jk1} It was argued that the transition between the magnetically 
ordered state for $K/J \alt 8$ and a striped (or valence-bond-solid, VBS) 
phase at 
higher $K/J$ is a continuous quantum phase transition, contrary to general 
expectations for an order-order transition. Subsequently, this transition 
was proposed to possibly be a realization of a ``deconfined'' quantum-critical
point.\cite{senthil} We have used the SSE algorithms to further study the 
quantum-critical scaling and finite-$T$ transitions in this model. However, 
in this paper we only briefly summarize the results and focus on the 
algorithmic issues. A full account of the results will be presented 
elsewhere.\cite{respaper}

For $K=0$, the J-K model reduces to the standard XY-model, which undergoes
a Kosterlitz-Thouless transition at $T_{\rm KT}/J \approx 
0.69$.\cite{loh,harada,XYrad} 
In the boson language, the system is a superfluid below $T_{\rm KT}$. 
The main features of the $T=0$ phase diagram for $K/J>0$ were presented in 
Ref.~\onlinecite{jk1}. Our most recent simulations \cite{jk1,respaper} show 
that the superfluid density vanishes at $K/J \approx 7.91$. At the same point, 
within the accuracy of our calculations, the ground state develops a stripe 
order, where the bond and plaquette strengths $\langle B_{ij}\rangle$ and 
$\langle P_{ijkl}\rangle$ are modulated at wave-vector ${\bf q}=(\pi,0)$ or 
$(0,\pi)$. This state can also be considered a columnar VBS, since not
all the bonds within the ``ladders'' of strong bonds are equal---the strongest
ones are those on the rungs of the ladders. The VBS order vanishes
at $K/J \approx 14.5$, in a first-order transition to an Ising-type 
antiferromagnetic state (a charge-density-wave, CDW, at ${\bf q}=
(\pi,\pi)$ in the boson picture). We have not observed any signs of
first-order behavior at the superfluid-VBS transition, nor any region of 
coexistence of the two phases. Numerically we can of course never
exclude an extremely weakly first-order transition or a very narrow
coexistence region. At the transition, we do observe power-law scaling 
with nontrivial exponents for the superfluid density as well as for the 
order parameter corresponding to the VBS phase. We have also recently
studied the evolution of the VBS phase boundaries when the system is 
coupled to an external magnetic field.\cite{melko}

The outline of the rest of this paper is the following: 
In Sec.~II we describe the SSE algorithm
for the J-K Hamiltonian. Implementations of the SSE scheme for various 
spin \cite{directed,tising} and 1D fermion \cite{pinaki} models have been 
discussed at length in several recent papers, but since the four-particle term 
necessitates a more complex sampling scheme, with some important new features,
we describe our algorithm in detail here. We have constructed two types
of cluster updates for sampling the SSE configurations; a directed-loop
update as well as a ``multi-branch'' cluster update. The latter significantly
reduces the autocorrelation times for large $K/J$. In Sec.~II we also 
discuss estimators for several important physical quantities. We discuss 
autocorrelation functions in the different ordered phases in Sec.~\ref{AC}.
In Sec.~IV we discuss modifications of the algorithm when different types
of potential-energy terms are included in addition to the $J$ and $K$ terms.
We conclude with a brief discussion in Sec.~V.

\section{Stochastic series expansion}

The SSE method \cite{sse1,sse2,sse3,sse4} is an efficient and widely
applicable generalization of Handscomb's \cite{handscomb} power-series method
for the $S=1/2$ Heisenberg model. It has previously been used for several
models with two-body interactions, including the pure XY-model [$K=0$ in 
Eq.~(\ref{ham})].\cite{XYrad} As in world-line Monte Carlo,
\cite{worldline} loop-cluster algorithms \cite{evertz} can speed 
up SSE simulations very significantly.\cite{sse4} Recently, a framework was
devised for constructing and optimizing loop-type algorithms under very 
general conditions.\cite{directed} Here we will apply this {\it directed 
loop} scheme to SSE simulations including the four-spin term.  A loop-type 
algorithm cannot be constructed for the pure $K$-model [$J=0$ in 
Eq.~(\ref{ham})], however, and the loops are also inefficient when $J/K \ll 
1$. We therefore also develop a new type of {\it multi-branch} cluster update, 
that can be used in combination with the directed loops, and enables efficient 
simulations for any $J/K$. The multi-branch update bares some resemblance to,
but is more complex than, a ``quantum-cluster'' update recently developed for 
the transverse Ising model.\cite{tising} 

Below, we give a brief general summary of the SSE method. We then develop
the directed loop and multi-branch cluster updates for the J-K Hamiltonian 
and discuss the SSE estimators of several important physical quantities. We 
present some illustrative results for autocorrelation functions obtained
with and without the multi-branch cluster update before concluding with a 
discussion of the directed-loop equations for the Hamiltonian with various 
diagonal interaction included.

\subsection{General SSE formalism}

To construct the SSE representation of a quantum mechanical expectation 
value at temperature $T=1/\beta$;
\begin{equation}
\langle A \rangle = {1\over Z}{\rm Tr }\{ A {\rm e}^{-\beta H} \},~~~~ 
Z={\rm Tr }\{{\rm e}^{-\beta H} \}, 
\end{equation}
the Hamiltonian is first written as a sum of elementary interactions
\begin{equation}
H = - \sum\limits_{t}\sum\limits_{a} H_{t,a},
\label{hsum}
\end{equation}
where in a chosen basis $\{ |\alpha \rangle \}$ the operators satisfy
\begin{equation}
H_{t,a}|\alpha \rangle \sim |\alpha^\prime \rangle ,
\end{equation}
where $|\alpha \rangle$ and $|\alpha^\prime \rangle$ are both basis states. 
The indices $t$ and $a$ refer to the operator types (various kinetic and 
potential terms) and the lattice units over which the interactions are 
summed (bonds, plaquettes, etc.). A unit operator $H_{0,0} \equiv 1$ is also 
defined. Using the Taylor expansion of e$^{-\beta H}$ truncated at order $M$,
the partition function can then be written as \cite{sse1}
\begin{equation}
Z = \sum\limits_\alpha \sum_{S_M} {\beta^n(M-n)! \over M!} 
    \left \langle \alpha  \left | \prod_{i=1}^M H_{t_i,a_i} 
    \right | \alpha \right \rangle ,
\label{zm}
\end{equation}
where $S_M$ denotes a sequence of operator-indices;
\begin{equation}
S_M = [t_1,a_1],[t_2,a_2],\ldots ,[t_M,a_M],
\end{equation}
and $n$ denotes the number of non-$[0,0]$ elements in $S_M$ (i.e., the
actual expansion-order of the terms). The finite truncation $M$ and the use 
of a fill-in operator $H_{0,0}$ are not strictly necessary \cite{sse2} but 
simplify some aspects of the algorithm. $M$ can be adjusted 
during the equilibration of the simulation, so that it always exceeds the 
highest power $n$ reached; $M = An_{\rm max}$, where a suitable 
value for the factor is $A \approx 1.25$. This leads to $M \sim \beta N$, 
where $N$ is the system volume, and the remaining truncation error is 
completely negligible. The adjustment of $M$ has been discussed in 
more detail in Ref.~\onlinecite{sse3}.

Defining a normalized state $|\alpha (p)\rangle$ obtained by acting on
$|\alpha \rangle = |\alpha (0)\rangle$ with the first $p$ operators in
the product in Eq.~(\ref{zm}),
\begin{equation}
|\alpha (p)\rangle \sim \prod_{i=1}^p H_{t_i,a_i} |\alpha \rangle ,
\label{propagated}
\end{equation}
the requirement for a non-zero contribution to $Z$ is the propagation
periodicity $|\alpha (M) \rangle = |\alpha (0) \rangle$. This implies 
considerable constraints on the off-diagonal operators in the product, and 
clearly the vast majority of the terms are zero. In an efficient
SSE method, transitions $(\alpha,S_M) \to (\alpha^\prime , S_M^\prime)$ 
satisfying detailed balance should be attempted only within the subset of 
contributing configurations. Although the details of such sampling procedures
to some extent depend on the model under study, three different classes of 
updates are typically used. We here summarize these in general terms, before
turning to the implementation for the J-K model:

(i) The expansion order 
$n$ is changed in {\it diagonal updates}, where a fill-in unit operator
is replaced by a diagonal operator from the sum (\ref{hsum}), and vice 
versa, i.e., $H_{0,0} \leftrightarrow H_{d,a}$, where the type-index $d$ 
corresponds to a diagonal operator in the basis used. 

(ii) Off-diagonal operators 
cannot be added and removed one-by-one with the periodicity constraint 
$|\alpha (M)\rangle = |\alpha (0)\rangle$ maintained. Local updates involving
two simultaneously replaced operators can be used for this purpose.\cite{sse2}
However, much more efficient cluster-type updates, which may involve a large 
number of operators, can also be constructed.\cite{sse4,tising} Here the 
general strategy is to find a set of operators $\{ t_i,a_i \}$, such that a 
new valid configuration can be obtained by changing only the type-indices 
$t_i$. For the J-K Hamiltonian, we will discuss two such updates; directed 
loops and multi-branch clusters.

(iii) A third type of update is one that affects only the state 
$|\alpha \rangle$. This state, which is just one out of the whole cycle of 
propagated states $|\alpha (p)\rangle$, can change also in the updates (ii)
involving off-diagonal operators. However, at high temperatures many sites 
will frequently have no operators acting on them. The local states at these
sites will then not be affected by the off-diagonal updates. They can instead
be randomly modified as they do not affect the weight. Such state updates 
can improve the statistics at high temperatures but are often not required 
for the sampling to be ergodic.

\subsection{Plaquette operators}

Turning now to the J-K model, we use the standard $z$-component basis 
\begin{equation}
|\alpha \rangle = |\sigma^z_i,\ldots,\sigma^z_N\rangle,~~ \sigma^z_i = \pm 1
\end{equation}
where $S_i^z = 1/2 \sigma^z_i$,
on lattices with $N=L_x\times L_y$ sites (or $N$ plaquettes). Typically we
consider square lattices, $L_x=L_y$, but some results for rectangular, $L_x\not=L_y$, 
systems have also been discussed.\cite{jk1} It is convenient to 
express all interactions in the Hamiltonian (\ref{ham}) in terms of 
plaquette operators,
\begin{eqnarray}
H_{1,a} & = &  CI_{ijkl}, \nonumber \\
H_{2,a} & = & (J/2)B_{ij}I_{kl}, \nonumber \\
H_{3,a} & = & (J/2)B_{jk}I_{il}, \label{hab} \\
H_{4,a} & = & (J/2)B_{kl}I_{ij}, \nonumber \\
H_{5,a} & = & (J/2)B_{li}I_{jk}, \nonumber \\
H_{6,a} & = &  KP_{ijkl}, \nonumber
\end{eqnarray}
where $I_{ij}$ and $I_{ijkl}$ are unit operators associated with bonds and
plaquettes, respectively, and the indexing is defined in Fig.~\ref{fig:labels}.
Up to a constant $NC$, the Hamiltonian is then given by a sum 
(\ref{hsum}), where the type index $t=1,\ldots,6$, and $a$ is the plaquette 
index; $a=1,\ldots ,N$. As explained above, there is also a unit operator 
$H_{0,0}=1$, which is not part of the Hamiltonian but has been introduced only
as a fill-in element for augmenting the operator-index sequences of length
$n < M$ in the truncated partition function (Eq.~(\ref{zm})) to $M$.

\subsection{Diagonal update}

Because there are no diagonal operators in the original Hamiltonian 
(\ref{ham}), the constant operators $H_{1,a}$ have been added in order to 
enable diagonal updates of the form $[0,0] \leftrightarrow [1,a]$ in $S_M$. 
For all elements $[a_p,t_p]$ with $t_p =0,1$, such substitutions can be 
carried out sequentially for $p=1,\ldots,M$. In the $\rightarrow$ direction, 
the plaquette index $a$ is chosen randomly among $1,\ldots,N$. The Metropolis
acceptance probabilities are then\cite{sse3}
\begin{eqnarray}
P([0,0] \rightarrow [1,a]) & = & {NC\beta\over M-n}, \label{padd} \\
P([1,a] \rightarrow [0,0]) & = & {M-n+1 \over NC\beta},\label{pdel} 
\end{eqnarray}
where $P >1$ should be interpreted as probability one. If an attempt to remove
a plaquette operator, i.e., $[1,a] \rightarrow [0,0]$, is not accepted, a new
plaquette index $a$ can be generated at random. Note that for this model, 
where the only diagonal operators are the added constants $H_{1,a}$, it is 
not necessary to keep track of the propagated states during the diagonal 
update. In general, e.g., if a diagonal interaction is added to the 
Hamiltonian (\ref{ham}), the constant $C$ in Eqs.~(\ref{padd}) and 
(\ref{pdel}) should be replaced by the matrix element 
$\langle \alpha (p)| H_{1,a_p} |\alpha (p-1) \rangle = 
\langle \alpha (p)| H_{1,a_p} |\alpha (p) \rangle$ of the diagonal operator 
in the propagated state at which the replacement is done.

\subsection{Linked vertices}

\begin{figure}
\includegraphics[width=3.5cm]{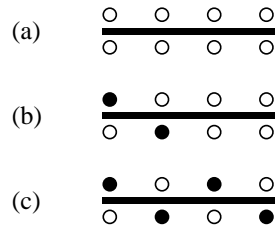}
\caption{Examples of vertices for the J-K model.
The solid and open circles correspond to up and down spins, respectively, 
before (beneath the bar) and after (above the bar) an operator has acted. 
(a) is one out of 16 diagonal C-vertices, (b) is one out of 32 J-vertices
(which flip two spins), and (c) one of the two K-vertices (which flip all
four spins).}
\label{fig:vertices}
\end{figure}

\begin{figure}
\includegraphics[width=6.5cm]{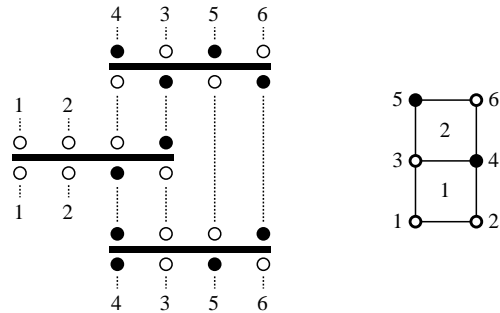}
\caption{The linked-vertex representation corresponding to the matrix
element $\langle \alpha |H_{6,2}H_{4,1}H_{4,2}|\alpha\rangle$ (left), where
the basis state $|\alpha \rangle = \vert\dn \dn \dn \up \up \dn\rangle$
(right). The bidirectional links are represented as dashed lines. The site 
and plaquette numbering for the six-site lattice is shown to the right. The 
numbers at the vertex legs indicate links across the periodic propagation 
boundary, and the corresponding spins are hence the same as in the state 
$|\alpha\rangle$. The numbers here also correspond to the site numbering of 
the lattice shown to the right.}
\label{fig:linked}
\end{figure}

In the directed loop and multi-branch cluster updates, which we will discuss 
below, it is useful to represent the matrix elements in Eq.~(\ref{zm}) as a
linked lists of ``vertices''.\cite{sse4} The weight of a configuration 
$(\alpha,S_M)$ can be written as
\begin{equation}
W(\alpha,S_M) = {\beta^n(M-n)! \over M!} \prod_{p=1}^M  W(p),
\label{wasm}
\end{equation}
where $W(p)$ is a {\it vertex weight}, which is simply the matrix element of
the corresponding plaquette operator at position $p$ in $S_M$;
\begin{equation}
W(p) = \langle \alpha(p) |H_{t_p,a_p}|\alpha (p-1)\rangle ,
\label{wp}
\end{equation}
which with the  operators (\ref{hab}) can take the values $C$, $J/2$, or
$K$. Since the loop and cluster updates are carried out within sectors of fixed
$n$ (only the diagonal update changes $n$), the fill-in operators $H_{0,0}$
are not needed in the linked-vertex representation. A vertex represents the 
local four-spin states on plaquette $a_p$ in the matrix element (\ref{wp}) 
before and after the plaquette operator has acted. These eight spin states
constitute the {\it legs} of the vertex. For the J-K model, there are three
classes of vertices, as illustrated in Fig.~\ref{fig:vertices}. The constant 
operators $H_{1,a}$ correspond to C-vertices (with weight $C$), the 
bond-flip operators $H_{2,a}$-$H_{5,a}$ to J-vertices (with weight $J/2$) 
and the plaquette-flip operators $H_{6,a}$ to K-vertices (with weight $K$).
An example of a linked-vertex representation of a term with three plaquette 
operators is shown in  Fig.~\ref{fig:linked}. The links connect vertex-legs 
on the same site, so that from each leg of each vertex, one can reach the 
next or previous vertex-leg on the same site (i.e., the links are 
bidirectional). In cases where there is only one operator acting on a given 
site, the corresponding ``before'' and ``after'' legs of the same vertex 
are linked to each other (as is the case with the legs on sites 1 and 2 
in Fig.~\ref{fig:linked}). 

During the simulation, the spin state $|\alpha \rangle$ and the operator list
$S_M$ are stored at all times. The linked-vertex representation is created
after each full sweep of diagonal updates. After the directed loop and
multi-branch cluster updates have been carried out, the changes are mapped
back into a new $|\alpha \rangle$ and $S_M$. We will not discuss here how 
these data structures are implemented and used in practice in a computer
program. The procedures are completely analogous to simulations with
two-body interactions, for which an implementation was described in detail in
 Ref.~\onlinecite{directed}.

\subsection{Directed loops}

In the original QMC loop algorithm,\cite{evertz1} spins are flipped along a 
one-dimensional closed path (the loop) on 
the space-time lattice of the discretized 
(Trotter-decomposed) or continuous \cite{beard} path-integral representation.
\cite{evertz} The path is self-avoiding, and a configuration can be subdivided
into loops that may be flipped independently of each other. Allowing the path
to self-intersect and backtrack, one can construct valid algorithms for a much
larger class of models. Such general loop-type algorithms have been 
constructed both for  continuous-time world-lines (the worm algorithm 
\cite{prokofev}) and for SSE (the operator-loop algorithm \cite{sse4}). The 
detailed balance equations---the directed loop equations---that must be 
satisfied when constructing general self-intersecting and back-tracking loops 
were recently derived within the SSE framework, and a generalization to the 
path integral 
representation was also shown.\cite{directed} Here we will implement the
directed-loop scheme for SSE sampling of the $J$ and $K$ terms.

In an SSE operator-loop algorithm, where the loops constitute connected 
strings of operators (or vertices in the linked-vertex representation),
\cite{sse4} the building of a loop consists of a series of steps, in each of
which a vertex is entered at one leg (the entrance leg) and an exit leg is 
chosen according to probabilities that depend on the entrance leg and the spin 
states at all the legs. The entrance to the following vertex is given by the 
link from the chosen exit leg. The spins at all vertex-legs visited are 
flipped during the loop building. 

The original starting point of the loop is chosen at random. Two 
{\it link-discontinuities} are created when the first pair of entrance and exit 
spins is flipped, i.e., the legs to which these are linked will be in 
different spin states (this is analogous to introducing the two sources in 
the worm algorithm \cite{prokofev}). Configurations contributing to $Z$ only 
contain links between legs in the same spin states. One of the discontinuities
will be propagated during the loop-building, whereas the other one will
remain at the original starting point. The loop closes when the propagating 
discontinuity reaches the stationary one, so that they annihilate each other. 
A new contributing configuration has then been generated. If the path is 
self-intersecting (which is not always the case \cite{sse4}), the changes 
in the configuration may in effect correspond to several disconnected loops.

\begin{figure}
\includegraphics[width=7cm]{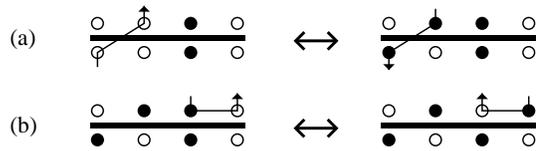}
\caption{Two examples of vertex-paths that are related in the directed loop 
scheme. (a) shows a process, and its reverse, where a C-vertex is 
transformed into a J-vertex. (b) shows two related J$\leftrightarrow$K
transformations. In the loop construction the spin states at the entrance 
and exit legs are flipped. The spin states shown in the vertices here are 
those before the flips have been carried out.}
\label{fig:related}
\end{figure}

When a vertex has been entered at a given leg,
the probabilities for choosing one out of the possible exit legs have 
to be chosen so that detailed balance is satisfied. In general, 
these probabilities are not unique, and in most cases the most 
evident ones involve high probabilities for {\it bounces}, where the exit and 
entrance legs are the same and the loop building hence backtracks one step.
\cite{sse4} It is normally\cite{Alet} desirable to minimize the probability of 
bounces. 
The directed loop scheme \cite{directed} systematizes the search for valid sets
of exit probabilities and enables a minimization of the bounce probability.
To construct the directed loop equations for the exit probabilities, weights 
are first assigned to all possible paths through a vertex from a given 
entrance leg. The sum of all these path weights must equal the {\it bare 
vertex weight} (\ref{wp}), i.e., the matrix element before the entrance and 
exits spins have been flipped. The actual normalized exit probability is 
the path weight divided by the bare vertex weight. The key element of the 
scheme is that weights for vertex-paths that constitute each other's reverses 
have to be equal in order for detailed balance to be fulfilled (a generalized
scheme where this is not necessarily the case has also been 
discussed recently \cite{Alet}). Examples of such related vertex-paths in the J-K 
model are shown in Fig.~\ref{fig:related}. The directed loop equations written down 
on the basis of these simple rules often form several different closed sets that can be 
solved for the path weights independently of each other. Because of 
symmetries, many of the equation sets can also be also identical. In general, 
the directed loop equations have an infinite number of solutions, which can 
be significantly restricted by minimizing the bounce probabilities. In some 
cases there is a unique minimum-bounce solution (sometimes with zero bounce 
probability), but often there is still a high degree of freedom left.
\cite{directed,harada2,olav}

\begin{figure}
\includegraphics[width=8cm]{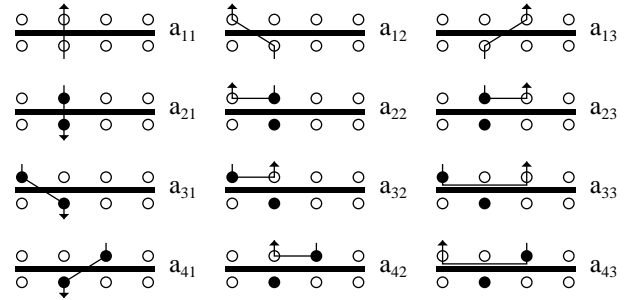}
\caption{A closed set of C and J vertex-paths, with their corresponding
weights $a_{ij}$ that have to satisfy the directed loop equations.}
\label{fig:xyvert}
\end{figure}

For the J-K model, a one dimensional path segment can in one step transform
a C-vertex into a J-vertex, and vice versa, an example of which is shown 
in Fig.~\ref{fig:related}(a). A J-vertex can be transformed into a 
K-vertex, and vice versa, as shown in Fig.~\ref{fig:related}(b). C- and 
K-vertices cannot be directly transformed into each other, however. As a 
consequence, the closed sets of vertex-paths that contain C$\leftrightarrow$J 
transformations are independent from those containing J$\leftrightarrow$K 
transformations. 

The closed sets containing C$\leftrightarrow$J transformations are similar to 
those for the XY-model,\cite{directed} although the sets are larger because a 
C-vertex can be transformed into two different J-vertices. As in the XY-model,
no bounces are required for detailed balance in this case, until we discuss
the inclusion of additional diagonal interactions in Sec.~IV. 
One closed set 
with C$\leftrightarrow$J transformations is shown in Fig.~\ref{fig:xyvert}. 
To construct such a set, one first selects a ``reference'' vertex (any vertex) 
and an entrance leg, and then finds all paths that 
lead to new valid vertices, sampling all allowed exit legs. 
This corresponds to the
first row of Fig.~\ref{fig:xyvert}, where the bounce process has not been 
included since, as will be shown below, its weight can be set to zero in this
case.
Each of the resulting vertices (i.e., when the entrance and exit
spins have been flipped) are then considered in turn, using as the entrance 
legs the exit legs from the previous step. This leads to rows two to four in 
Fig.~\ref{fig:xyvert}. The procedure is repeated for each new combination of 
vertex and exit leg that is created. This systematically generates all pairs 
of vertex-paths that constitute each other's reverses, i.e., those that must 
have equal weights for detailed balance to be satisfied. In the case 
considered here, no 
new vertex-paths are created after row four, as the reverse of each path has
then already been generated. The set is hence closed. Other closed sets are 
constructed by picking a starting vertex and entrance leg combination that 
has not yet appeared within the sets already completed. This is repeated
until all vertices and entrance legs combinations have been exhausted.

The directed loop equations corresponding to the closed set shown in 
Fig.~\ref{fig:xyvert} are
\begin{eqnarray}
a_{11} + a_{12} + a_{13}  & = & W_1 = C,  \nonumber \\
a_{21} + a_{22} + a_{23}  & = & W_2 = C,  \label{eqset1} \\
a_{31} + a_{32} + a_{33}  & = & W_3 = J/2,  \nonumber \\
a_{41} + a_{42} + a_{43}  & = & W_4 = J/2,  \nonumber 
\end{eqnarray}
where the weights $a_{ij}$ are identified with the paths in the figure
and $W_i$ are the bare vertex weights before the entrance and exit spins 
have been flipped. Detailed balance requires that the weights corresponding
to opposite vertex-paths are equal, i.e.,
\begin{eqnarray}
\label{balancereq}
a_{21} & = & a_{11}, \nonumber \\
a_{31} & = & a_{12}, \nonumber \\
a_{32} & = & a_{22}, \nonumber \\
a_{41} & = & a_{13}, \\
a_{42} & = & a_{23}, \nonumber \\
a_{43} & = & a_{33}. \nonumber
\end{eqnarray}
The weights also have to be positive definite, since they are related to 
probabilities by dividing with the positive matrix elements $W_i$.
Even with these constraints, 
the solution is not unique. One can reasonably assume that the most efficient 
solution also has equal weights for paths that are related by symmetries, 
e.g., $a_{12}=a_{13}$. Using all such symmetries, the solution is still not 
unique, however. It can be expected that it is efficient to maximize the 
weights of the paths that transform a C-vertex into a J-vertex, which is
equivalent to minimizing the weights of the {\it continue-straight} paths that 
transform a C-vertex into another C-vertex. We have no proof of our assertion
that this is a good strategy, but as it is a quite challenging task to
investigate all possible valid solutions, we will use it and leave other 
possibilities for future studies (this issue has in fact recently been
addressed in the context of other models\cite{PRHH}). In Fig.~\ref{fig:xyvert}, 
there are only two C$\to$C paths; the pair with weights $a_{11}$, $a_{21}$. 
The minimum value of these is $a_{11}=a_{21}=C-J/2$, which also implies 
$C \ge J/2$. There are now enough conditions to render a unique solution 
to this set of directed loop equations;
\begin{equation}
\begin{array}{ll}
a_{11} & = C-J/2, \\
a_{21} & = C-J/2, \\
a_{31} & = J/4, \\
a_{41} & = J/4,
\end{array}
~~
\begin{array}{ll}
a_{12} & = J/4, \\
a_{22} & = J/4, \\
a_{32} & = J/4, \\
a_{42} & = J/4,
\end{array}
~~
\begin{array}{ll}
a_{13} & = J/4, \\
a_{23} & = J/4, \\
a_{33} & = 0, \\
a_{43} & = 0.
\end{array}
\label{cjsolution}
\end{equation}
The actual exit probabilities $P^a_{ij}=a_{ij}/W_i$ are
\begin{equation}
\begin{array}{ll}
P^a_{11} & = 1-J/2C, \\
P^a_{21} & = 1-J/2C, \\
P^a_{21} & = 1/2, \\
P^a_{41} & = 1/2,
\end{array}
~
\begin{array}{ll}
P^a_{12} & = J/4C, \\
P^a_{22} & = J/4C, \\
P^a_{22} & = 1/2, \\
P^a_{42} & = 1/2,
\end{array}
~
\begin{array}{ll}
P^a_{13} & = J/4C, \\
P^a_{23} & = J/4C, \\
P^a_{23} & = 0, \\
P^a_{43} & = 0,
\end{array}
\end{equation}
where the superscript $a$ is used as a reminder that these probabilities
correspond to the paths shown in Fig.~\ref{fig:xyvert}. Note 
that the probabilities here depend only on the type of vertex transformation,
C$\to$C ($P=1-J/2C$), C$\to$J ($P=J/2C$), J$\to$J ($P=0$), or J$\to$C 
($P=1/2$), which can aid the implementation of the probability tables in the code.
All other sets with C$\leftrightarrow$J transformations are 
either related by trivial symmetries to that shown in Fig.~\ref{fig:xyvert}
or are very similar to it. The exit probabilities are given simply by the 
type of the corresponding vertex transformation exactly as above.

The directed loop equations for the closed sets of paths that involve
J$\leftrightarrow$K transformations sometimes require non-zero bounce 
probabilities. A closed set of paths is shown in Fig.~\ref{fig:kjvert}. 
The corresponding equations for the path weights $b_{ij}$ are
\begin{eqnarray}
b_{11} + b_{12} + b_{13} + b_{14} + b_{15}  & = & J/2,  \nonumber \\
b_{21} + b_{22} + b_{23} + b_{24} + b_{25}  & = & J/2,  \nonumber \\
b_{31} + b_{32} + b_{33} + b_{34} + b_{35}  & = & K,   \\
b_{41} + b_{42} + b_{43} + b_{44} + b_{45}  & = & J/2,  \nonumber \\
b_{51} + b_{52} + b_{53} + b_{54} + b_{55}  & = & J/2.  \nonumber 
\end{eqnarray}
Again, it is in general advantageous to minimize the bounce probabilities, i.e.,
the bounce weights $b_{i,5}$ above. For $K \le 2J$ all the bounce weights can
in fact be zero. The weight of the continue-straight paths (e.g., $b_{11}$), 
which here transform a J-vertex into a J-vertex with the same spin flips
(i.e., the same plaquette operator), can be set to zero. A symmetric 
$K \le 2J$ solution is then:
\begin{widetext}
\begin{equation}
\begin{array}{ll}
b_{11} & = 0, \\
b_{21} & = 0, \\
b_{31} & = K/4, \\
b_{41} & = J/4-K/8 \\
b_{51} & = J/4-K/8,
\end{array}
~~~
\begin{array}{ll}
b_{12} & = K/4, \\
b_{22} & = K/4, \\
b_{32} & = K/4, \\
b_{42} & = 0, \\
b_{52} & = 0
\end{array}
~~~
\begin{array}{ll}
b_{13} & = J/4-K/8, \\
b_{23} & = J/4-K/8, \\
b_{33} & = K/4, \\
b_{43} & = J/4-K/8, \\
b_{53} & = K/4, 
\end{array}
~~~
\begin{array}{ll}
b_{14} & = J/4-K/8, \\
b_{24} & = J/4-K/8, \\
b_{34} & = K/4, \\
b_{44} & = K/4, \\
b_{54} & = J/4-K/8,
\end{array}
~~~
\begin{array}{ll}
b_{15} & = 0, \\
b_{25} & = 0, \\
b_{35} & = 0, \\
b_{45} & = 0, \\
b_{55} & = 0.
\end{array}
\label{bij1}
\end{equation}
For $K>2J$, the bounce weight $b_{35}$ has to be non-zero for a 
positive-definite solution. Minimizing this weight one obtains the
following solution:
\begin{equation}
\begin{array}{ll}
b_{11} & = 0, \\
b_{21} & = 0, \\
b_{31} & = J/2, \\
b_{41} & = 0 \\
b_{51} & = 0,
\end{array}
~~~
\begin{array}{ll}
b_{12} & = J/2, \\
b_{22} & = J/2, \\
b_{32} & = J/2, \\
b_{42} & = 0, \\
b_{52} & = 0
\end{array}
~~~
\begin{array}{ll}
b_{13} & = 0, \\
b_{23} & = 0, \\
b_{33} & = J/2, \\
b_{43} & = 0, \\
b_{53} & = J/2, 
\end{array}
~~~
\begin{array}{ll}
b_{14} & = 0, \\
b_{24} & = 0, \\
b_{34} & = J/2, \\
b_{44} & = J/2, \\
b_{54} & = 0,
\end{array}
~~~
\begin{array}{ll}
b_{15} & = 0, \\
b_{25} & = 0, \\
b_{35} & = K-2J, \\
b_{45} & = 0, \\
b_{55} & = 0.
\end{array}
\label{bij2}
\end{equation}
\begin{figure*}
\includegraphics[width=14cm]{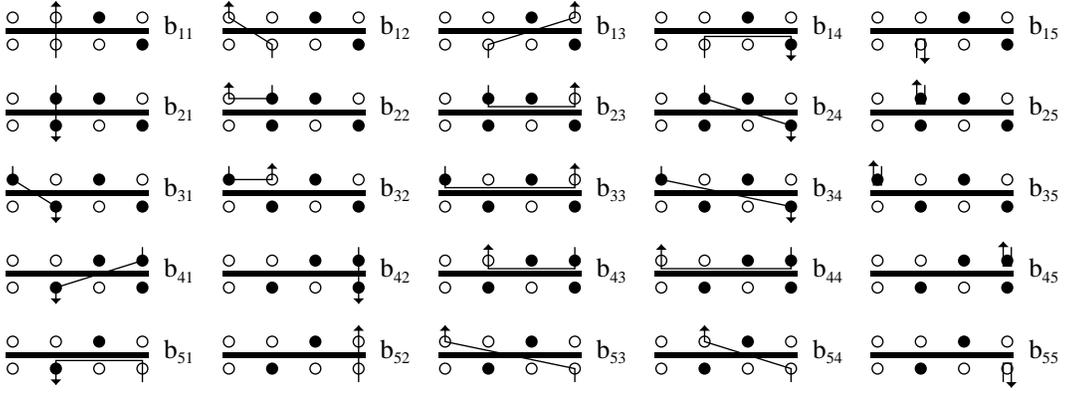}
\caption{A closed set of vertex-paths with J$\leftrightarrow$K
transformations, labeled by their path weights $b_{ij}$.}
\label{fig:kjvert}
\end{figure*}
The exit probabilities are hence, for $K \le 2J$:
\begin{equation}
\begin{array}{ll}
P^b_{11} & = 0, \\
P^b_{21} & = 0, \\
P^b_{31} & = 1/4, \\
P^b_{41} & = 1/2-K/4J \\
P^b_{51} & = 1/2-K/4J,
\end{array}
~~~
\begin{array}{ll}
P^b_{12} & = K/2J, \\
P^b_{22} & = K/2J, \\
P^b_{32} & = 1/4, \\
P^b_{42} & = 0, \\
P^b_{52} & = 0
\end{array}
~~~
\begin{array}{ll}
P^b_{13} & = 1/2-K/4J, \\
P^b_{23} & = 1/2-K/4J, \\
P^b_{33} & = 1/4, \\
P^b_{43} & = 1/2-K/4J, \\
P^b_{53} & = K/2J, 
\end{array}
~~~
\begin{array}{ll}
P^b_{14} & = 1/2-K/4J, \\
P^b_{24} & = 1/2-K/4J, \\
P^b_{34} & = 1/4, \\
P^b_{44} & = K/2J, \\
P^b_{54} & = 1/2-K/4J,
\end{array}
~~~
\begin{array}{ll}
P^b_{15} & = 0, \\
P^b_{25} & = 0, \\
P^b_{35} & = 0, \\
P^b_{45} & = 0, \\
P^b_{55} & = 0,
\end{array}
\label{pbij1}
\end{equation}
and for $K > 2J$:
\begin{equation}
\begin{array}{ll}
P^b_{11} & = 0, \\
P^b_{21} & = 0, \\
P^b_{31} & = J/2K, \\
P^b_{41} & = 0 \\
P^b_{51} & = 0,
\end{array}
~~~
\begin{array}{ll}
P^b_{12} & = 1, \\
P^b_{22} & = 1, \\
P^b_{32} & = J/2K, \\
P^b_{42} & = 0, \\
P^b_{52} & = 0
\end{array}
~~~
\begin{array}{ll}
P^b_{13} & = 0, \\
P^b_{23} & = 0, \\
P^b_{33} & = J/2K, \\
P^b_{43} & = 0, \\
P^b_{53} & = 1, 
\end{array}
~~~
\begin{array}{ll}
P^b_{14} & = 0, \\
P^b_{24} & = 0, \\
P^b_{34} & = J/2K, \\
P^b_{44} & = 1, \\
P^b_{54} & = 0,
\end{array}
~~~
\begin{array}{ll}
P^b_{15} & = 0, \\
P^b_{25} & = 0, \\
P^b_{35} & = 1-2J/K, \\
P^b_{45} & = 0, \\
P^b_{55} & = 0.
\end{array}
\label{pbij2}
\end{equation}
Note that the solution is continuous across $K=2J$.

~~~~

\end{widetext}

Also in this case the probabilities are seen to depend only on the type
of vertex class transformation, J$\to$J, J$\to$J', J$\to$K, K$\to$J, or
K$\to$K (bounce). Here one has to distinguish between a continue-straight
J$\to$J transformation where the spin-flip remains on the same bond (e.g.,
$b_{11}$), and a J$\to$J' transformations where the spin-flip moves to a 
neighboring bond on the plaquette (e.g., $b_{13}$).

\begin{figure}
\includegraphics[width=2.5cm]{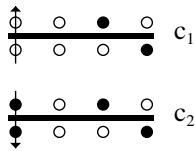}
\caption{A closed set of two vertex-paths, where only the continue-straight
process is allowed and a J-vertex is transformed into another
J-vertex with probability 1.}
\label{fig:jjvert}
\end{figure}

There is one more type of closed set of vertex-paths, an example of which 
is shown in Fig.~\ref{fig:jjvert}. In this case, neither a valid K-vertex 
nor a J-vertex with the flip moved to a different nearest-neighbor pair 
can be reached from the J-vertex and the chosen entrance leg. As the two
vertices shown have the same bare weights, no bounce processes have to
be included and the exit is unique:
\begin{eqnarray}
P^c_{1} & = 1, \nonumber \\
P^c_{2} & = 1.
\label{pjj}
\end{eqnarray}

All closed sets of vertex-paths can be related to those shown in 
Figs.~\ref{fig:xyvert}, \ref{fig:kjvert}, and \ref{fig:jjvert}, and in all 
cases the probabilities depend only on how the paths transform the vertices
between the classes C, J, and K. This simplifying property will be discussed
further in Sec.~IV, where we consider inclusions of additional diagonal 
interactions in the Hamiltonian.  In that case the solutions to the 
equations become more complicated, however the directed-loop framework is still
required in order to develop efficient codes.  For the case of zero diagonal
interactions, Eq.~(\ref{ham}), the exit probabilities for the J-K model are 
summarized in Table \ref{probs}.

\begin{table}
\caption{\label{probs}
All exit probabilities for the J-K model. The initial vertex class is
indicated in front of the square bracket, and the new class after the arrow.
All the possible vertex classes that can be generated from a given vertex 
and entrance leg are listed within the square bracket. In cases where more 
than one vertex of a given class can be generated, the corresponding symbol 
appears multiple times. $J$, $J'$, $J''$ denote subclasses of J-vertices 
in which different spin pairs are flipped. The sets $a,b,c$ correspond
to Figs.~\ref{fig:xyvert}, \ref{fig:kjvert}, \ref{fig:jjvert}; all other
sets are related to these by symmetries. The only bounce process is 
K$\to$K; all C$\to$C and J$\to$J cases correspond to continue-straight 
paths.}
\begin{ruledtabular}
\begin{tabular}{llll}
Vertex transformation          & set & $P(K \le 2J)$   &  $P(K > 2J)$   \\  
\hline
\\ 
C-[C,J,J']$\to$ C              & a & $1-J/2C$        &  $1-J/2C$       \\  
C-[C,J,J']$\to$ J,J'           & a & $J/4C$          &  $J/4C$        \\  
J-[C,C,J']$\to$ C,C            & a & $1/2$           &  $1/2$         \\  
J-[C,C,J']$\to$ J'             & a & $0$             &  $0$           \\  
J-[J,J',J'',K]$\to$ J          & b & $0$             &  $0$           \\  
J-[J,J',J'',K]$\to$ J',J''     & b & $1/2-K/4J$      &  $0$           \\  
J-[J,J',J'',K]$\to$ K          & b & $K/2J$          &  $1$           \\  
K-[J,J,J',J',K]$\to$ J,J,J',J' & b & $1/4$           &  $J/2K$        \\  
K-[J,J,J',J',K]$\to$ K~        & b & $0$             &  $1-2J/K$      \\  
J-[J]$\to$ J                   & c & $1$             &  $1  $         \\  
\end{tabular}
\end{ruledtabular}
\end{table}

To carry out a directed loop update, a vertex-leg is first chosen at random.
This entrance leg together with all the spin states on the vertex determine 
which one of the exit probabilities in Table \ref{probs} should be applied 
when generating the exit leg. These probabilities can be stored in a
pre-generated table. When the exit has been selected, the link from it 
is used to enter another vertex, from which an exit is again chosen, etc.,
until the loop closes. The number of loops to be generated during each Monte 
Carlo step is adjusted such that the total number of vertices visited is, 
on average, of the same order as (e.g., equal to or twice) the number of 
vertex-legs ($8n$), e.g., $4\langle n\rangle$ or $4M$.

In some cases, a loop can become very long before it closes. In order to avoid
problems with loops that do not close within a reasonable time, one can impose
a maximum loop length. If this limit is exceeded, the loop building is 
terminated and the changes in the vertices are disregarded. This does not 
introduce any bias in quantities measured in the $(\alpha,S_M)$ 
representation. In practice, the termination can easily be accomplished
by simply exiting the loop-update routine without mapping 
the linked-vertex representation back into a state $|\alpha \rangle$ and 
an operator list $S_M$; in order to discard only the loop currently under
construction, its history would have to be stored. Hence, not only the 
terminated loop itself is discarded, but also all other loops constructed 
since the previous diagonal update. This is not a problem as long 
termination does not occur frequently. We typically set the maximum 
loop length to $\approx 100\langle n\rangle$, and the fraction of terminated 
loops is then very small.

\begin{figure}
\includegraphics[width=6.5cm]{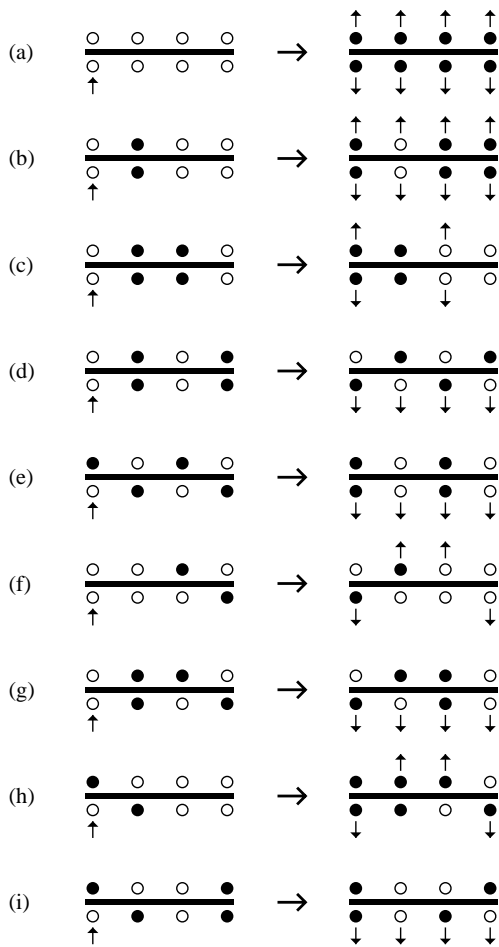}
\caption{Vertex transformations in the multi-branch cluster update.
The entrance leg is denoted by an arrow pointing into the vertices 
on the left. In the updated vertices to the right, the spins at the 
outgoing arrows have been flipped. The branching for all entrance legs 
and vertices not shown here are obtained by applying trivial symmetries 
to one of the cases shown in (a)-(i).}
\label{fig:mcluster}
\end{figure}

\begin{figure}
\includegraphics[width=6.5cm]{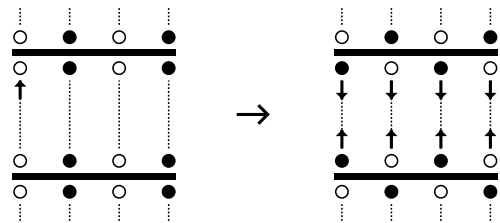}
\caption{Multi-branch cluster update in which two C-vertices are transformed
into two K-vertices. The initial entrance leg is at the inward pointing arrow
in the linked-vertex representation to the left. The resulting vertices
with their arrows indicating legs visited are shown to the right.}
\label{fig:mflipped}
\end{figure}

\subsection{Multi-branch clusters}

Since a K-vertex cannot be generated directly out of a C-vertex, but requires
the presence of J-vertices, the directed loop update cannot be used when 
$J=0$. As will be demonstrated in Sec.~\ref{AC}, it is also inefficient for 
large $K/J$. This can be understood from Table \ref{probs}, where the bounce 
probability off a K-vertex is seen to approach $1$ as $K\to \infty$. In 
principle the directed-loop update, in combination with the diagonal update, 
is ergodic for any finite $K/J$, but for $K/J \agt 12$ it becomes difficult
to obtain good results this way. In order to improve the performance for 
large $K/J$, a type of multi-branch cluster update is developed here. It is
similar to a quantum-cluster update recently developed for the transverse
Ising model,\cite{tising} 
where it can be considered a direct generalization of the
classical Swendsen-Wang algorithm.\cite{swalg} The multi-branch cluster 
update for the J-K model is more complex, due to the larger number of
different interaction vertices and the multitude of possible transformations
among them.

In order to transform a C-vertex directly into a K-vertex, spins at four 
legs have to be flipped. If this is done, spins also have to be flipped
at all the legs to which these four legs are linked. This will in turn
force additional spin flips in the vertices to which they are linked, etc. 
Clearly, such a process can branch out very quickly to a large number of 
vertices. Even if a scheme can be found where detailed balance is maintained,
there is in general nothing that guarantees that the process ever terminates 
(i.e., the cluster may not complete). For the J-K model, this proliferation
problem can be solved by defining a unique set of exit legs, given an entrance
leg and all the eight spin states of the vertex. If the constant $C$ is chosen
equal to $K$, which with the directed loop probabilities in Table \ref{probs} 
can be done if $K \ge J/2$, C$\leftrightarrow$K transformations will lead to 
no weight changes. If J-vertices are always transformed into other J-vertices, 
there are also no weight changes in these processes. A constructed cluster 
can therefore be flipped with probability one. One can also subdivide the 
whole linked vertex list into clusters that can be flipped independently of 
each other with probability $1/2$. This Swendsen-Wang-type approach will be 
used here.

Fig.~\ref{fig:mcluster} shows the branching rules for all different types of 
vertices. The cases (d) and (e) correspond to C$\leftrightarrow$K 
transformations. In all other cases the vertex class does not change, but note
that J-vertices are transformed into J-vertices with a different pair of 
flipped spins (i.e., the corresponding plaquette operator changes). 
The outgoing arrows point to entrances to other vertices, to 
which the same branching rules are applied. However, if an exit leg is linked
to a leg which has already been visited, this leg should not be visited again.
In terms of the graphical representation used in Fig.~\ref{fig:mcluster}, a 
vertex-leg should not be entered if it already has an outgoing arrow. If a 
vertex is entered for a second time, and hence has arrows at four legs 
(those with eight exit legs in Fig.~\ref{fig:mcluster} can clearly only
be visited once), the second set of exit legs are exactly the four that were 
not previously assigned arrows. In other words, all vertices in (c)-(i) can be
assigned outgoing arrows in two different ways, and the set chosen is the
one to which the entrance leg belongs. Furthermore, the two sets of mutually 
exclusive exit legs are exactly the same in the vertices obtained when the 
legs in one of these sets are flipped. This solves the proliferation problem, 
since it is guaranteed that a vertex-leg can be visited only once. It
also allows for independent flips of all clusters.

To start a cluster, a vertex-leg which does not belong to a cluster already 
constructed is first chosen at random, and the branching is assigned according
to the rules defined in Fig.~\ref{fig:mcluster}. Flags are set on all 
the exit legs, to indicate that they have been visited (corresponding to the 
outgoing arrows Fig.~\ref{fig:mcluster}). Note that the entrance also becomes
an exit leg with an outgoing arrow. If the cluster is to be flipped (which it
should with probability 1/2), the spins at all the exit legs are flipped. 
All exit legs are put on a stack. They are subsequently picked one-by-one 
from the stack, and the legs to which they are linked are used as entrance 
legs to other vertices if they have not yet been visited, i.e., 
these legs are flipped and put on the stack only if they have not been visited 
before. In the graphical representation, a cluster-branch ends when an arrow 
is encountered. The whole cluster is completed when all arrows point to other
arrows; the stack with unprocessed entrance legs is then empty. A completed 
cluster with only two vertices is illustrated in Fig.~\ref{fig:mflipped}.

\begin{figure}
\includegraphics[width=7.5cm]{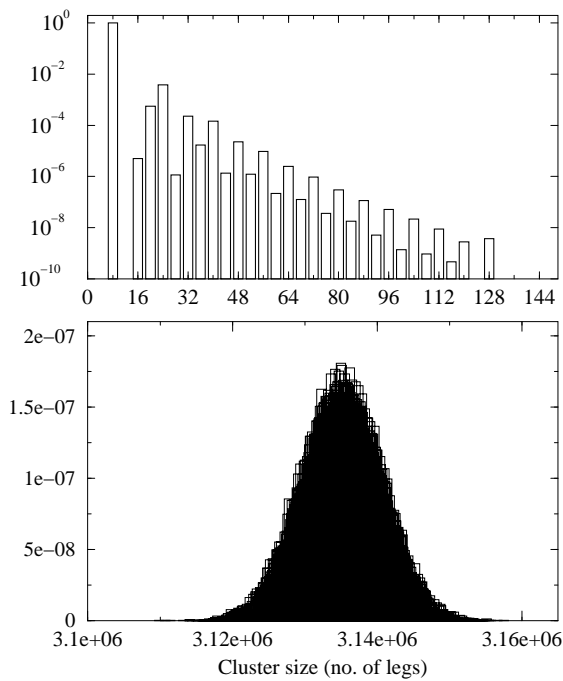}
\caption{Probability
histogram of cluster sizes (measured as the number of vertex legs in a
cluster) produced by multi-branch cluster updates of a 16$\times$16 lattice at
$K/J=80$ and $\beta=32$. The average length of the operator list for this
simulation was $\langle n \rangle \approx 4.35\times 10^5$ (i.e., approximately 
$3.5 \times 10^6$ vertex legs).  Data in the upper figure is for the 
smaller bins, while data in the lower figure was re-binned to 100 legs per bin.
All intermediate occupations were measured as zero.}
\label{clustHist}
\end{figure}
Although the autocorrelation measurements discussed in Sec.~\ref{AC}
provide a quantification of the ``effectiveness'' of the multi-branch cluster 
updates, we pause here to simply illustrate the cluster characteristics 
as implemented for our J-K model. A histogram of cluster sizes generated
at large $K/J$ is shown in Fig.~\ref{clustHist}.
Clearly, the vast majority of clusters built in this case are small eight-leg 
clusters (an example of which is illustrated in Fig.~\ref{fig:mflipped}), 
with significant occupation in the smaller bins up to clusters of size 128.
A second peak occurs in the histogram at much larger bin sizes, approaching 
the total number of operators. Clearly, the efficiency of the algorithm would 
be much greater if clusters occurred with sizes that were more evenly 
distributed 
between 8 and $8 n$, but in any case the multi-branch update
does significantly improve the performance for large $K/J$ (as will be shown
in Sec.~III). This can probably be explained by the fact that the 
directed-loop updates become very inefficient as $K/J \to \infty$, and hence the
multi-branch clusters help significantly even though the cluster-size 
distribution is not optimal.

\subsection{Physical observables}

In this section we summarize the physical observables relevant to studies
of the J-K model,\cite{roger3d,jk1,melko,respaper} and present the estimators 
used to evaluate them in the
SSE method. The general forms of the estimators have been derived in previous
papers;\cite{sse1,sse2,sse3,irsse} here we only apply those derived forms
to the particular quantities of interest for the J-K model.

We typically carry out measurements on the configurations generated after 
every Monte Carlo step (MCS), with an MCS defined as a sweep of diagonal 
updates, followed by construction of the linked vertex list, in which a 
fixed number of loop updates are carried out. In the same linked list, all 
multi-branch clusters are constructed and flipped with probability $1/2$.
After this, the updated vertex list is mapped back into a new state
$|\alpha \rangle$ and an operator list $S_M$. This is the representation
used for the measurements. The fill-in elements $H_{0,0}$ in $S_M$ are 
irrelevant at this stage, and we therefore now consider the reduced 
list $S_n$ without these operators. There are hence $n+1$ propagated states 
$|\alpha (p)\rangle=|\sigma^z_1(p),\ldots,\sigma^z_N(p)\rangle$, which 
are obtained one-by-one when operating with the first $p$ operators,
$p=0,\ldots,n$, on the initially stored state 
$|\alpha (0) \rangle = |\alpha (n) \rangle$. 
Although measurements can involve all the states, at any given time only 
a single $|\alpha (p)\rangle$ has to be stored.

The $z$-component of the spin-spin correlation function can be easily 
obtained, as it is diagonal in the representation used. Equal-time 
correlations can be averaged over the propagated states, i.e.,
\begin{equation}
\langle S^z_k S^z_l \rangle = {1\over 4}
\left \langle {1\over n} \sum\limits_{p=0}^{n-1} 
\sigma^z_k (p) \sigma^z_l (p) \right \rangle ,
\label{sumss}
\end{equation}
where in the special case $n=0$, which occurs in practice only for small 
$N$  at very high temperatures, the averaged sum should be replaced by 
$\sigma^z_k(0)\sigma^z_l(0)$. Since states $p$ and $p+1$ differ only by two 
or four flipped spins, the sum in (\ref{sumss}) can be replaced by a sum where
only, e.g., every $N^{\rm th}$ state is included. We often consider the Fourier 
transform of the correlation function, i.e., the static spin structure factor
\begin{equation}
S_s(q_x,q_y) = {1\over N}\sum\limits_{k,l} 
{\rm e}^{i({\bf r}_k-{\bf r}_l)\cdot {\bf q}}
\langle S^z_kS^z_l \rangle,
\end{equation}
where $r_i=(x_i,y_i)$ is the lattice coordinate (with lattice spacing $1$)
and $q=(q_xx,q_y)$ is the wave-vector. We also study the corresponding 
static susceptibility,
\begin{equation}
\chi_s(q_x,q_y) = {1\over N}\sum\limits_{k,l} 
{\rm e}^{i({\bf r}_k-{\bf r}_l)\cdot {\bf q}}
\int\limits_0^\beta \langle S^z_k(\tau)S^z_l (0) \rangle .
\label{chis}
\end{equation}
It has been shown \cite{sse1} that the SSE estimator for the Kubo integral is 
\begin{eqnarray}
& &\int\limits_0^\beta d\tau \langle S^z_k(\tau)S^z_l (0) \rangle = 
{\beta\over 4} \left\langle {1 \over n(n+1)} \times \right. \\  
& & \left. \left [ \left (\sum\limits_{p=0}^{n-1} \sigma^z_k (p) \right )
\left ( \sum\limits_{p=0}^{n-1} \sigma^z_l (p) \right ) +
\sum\limits_{p=0}^{n-1} \sigma^z_k (p) \sigma^z_l (p) \right ]\right \rangle .
\nonumber 
\label{kuboss}
\end{eqnarray}
Here the first term typically dominates; it is obtained by first summing the 
spins at $k$ and $l$ over the propagated states, and then multiplying the 
sums. The full sums must clearly be calculated here, but one can still take 
advantage of the fact that only two or four out of the $N$ spins $\sigma^z_k 
(p)$ change at every propagation $p \to p+1$. 
One can thus evaluate the sums for all sites
$k$ in $\sim n \sim N\beta$ steps. The second term in (\ref{kuboss}) vanishes 
as $N\to \infty$, but typically it gives a non-negligible relative 
contribution for small $N$ calculations and should always be kept. 
This sum is the same as in the equal-time correlation 
(\ref{sumss}) and can again be replaced by a partial summation without 
introducing a bias. In the case $n=0$, the whole expression within 
$\langle\rangle$ in Eq.~(\ref{kuboss}) should be replaced by 
$\sigma^z_k (0)\sigma^z_l (p)$.

We are also interested in the spin stiffness, or the superfluid density in the
boson representation, which at $T=0$ is defined by
\begin{equation}
\rho_s = {\partial ^2 E(\phi) \over \partial\phi ^2},
\label{rhos}
\end{equation}
where $E(\phi) = \langle H(\phi)\rangle /L^2$ is the ground state energy
per site and $\phi$ is a twist which is imposed on all bonds ($i,j$) in either 
the $x$ or $y$ lattice direction, so that the corresponding bond operators 
(\ref{bond}) become
\begin{eqnarray}
B_{ij}(\phi) & =  &
\cos{(\phi)} (S^x_iS^x_j  + S^y_iS^y_j) \nonumber \\
&& + \sin{(\phi)} (S^x_iS^y_j  - S^y_iS^x_j).
\end{eqnarray}
This leads to a shift in the ground state energy $E$ to second order
in $\phi$. With the plaquette operator $P_{ijkl}(\phi)$, the leading-order
energy shift is $\propto \phi^4$, and hence it will not appear in the
estimator for the stiffness. The derivative at $\phi=0$ in Eq.~(\ref{rhos}) 
can therefore be directly estimated using the winding number fluctuations 
in the SSE simulations,\cite{sse3} in a way very similar to the way it is 
done in path integral methods.\cite{pollock} Defining the winding numbers 
$w_x$ and $w_y$ as
\begin{eqnarray}
w_\alpha = (N^+_\alpha - N^+_\alpha)/L,
\end{eqnarray}
where $N^\pm_\alpha$ denote the number of operators in $S_n$ which transport 
a boson (or spin-$\uparrow$) $\pm 1$ lattice steps in the $\alpha$-direction.
In the J-K model, only the bond operators $B_{ij}$ can transfer a net number
of particles; in terms of the corresponding plaquette operators (\ref{hab}),
the pairs $H_{2,a}$, $H_{4,a}$ and $H_{3,a}$, $H_{5,a}$ transfer particles 
along the $x$- and $y$-axis, respectively. By operating successively with
all operators in $S_n$ on the state $|\alpha\rangle$ one can determine all
the numbers $N^\pm_\alpha$ needed to obtain the winding numbers. The stiffness
is then given by
\begin{equation}
\rho_s = {1\over 2\beta}\langle w_x^2 + w_y^2 \rangle .
\label{rhow2}
\end{equation}
At finite $T$, the ground state energy $E(\phi)$ in Eq.~(\ref{rhos}) should
be replaced by the free energy $F(\phi)$. It turns out that this leads to
exactly the same estimator, Eq.~(\ref{rhow2}).  A detailed derivation of this
well known result \cite{pollock} for lattice models has been presented in 
Ref.~\onlinecite{cuccoli}.

In order to detect the modulations of the bond and plaquette expectation
values $\langle B_{ij}\rangle$ and $\langle P_{ijkl}\rangle$ in the striped
phase, one can use open boundary conditions in order to break the 
translational symmetry. In order to break the $90^\circ$ rotational symmetry,
rectangular lattices can be used. On these lattices one can observe a unique 
bond/plaquette pattern.\cite{jk1} However, for careful finite-size scaling 
studies it is preferable to consider periodic $L\times L$ lattices, on which
all bond and plaquette expectations average to uniform values. We hence instead
consider the corresponding correlation functions, and also calculate the
associated susceptibilities. The static plaquette structure factor is
defined as
\begin{equation}
S_p(q_x,q_y) = {1\over N}\sum\limits_{a,b} 
{\rm e}^{i({\bf r}_a-{\bf r}_b)\cdot {\bf q}}
\langle P_aP_b \rangle,
\label{strp}
\end{equation}
where $P_a$ is the plaquette operator (\ref{plaquette}) with the plaquette 
subscript $a$ defined in Fig.~\ref{fig:labels}. The corresponding
susceptibility is completely analogous to Eq.~(\ref{chis}),
\begin{equation}
\chi_p(q_x,q_y) = {1\over N}\sum\limits_{a,b} 
{\rm e}^{i({\bf r}_a-{\bf r}_b)\cdot {\bf q}}
\int\limits_0^\beta d\tau \langle P_a(\tau)P_b (0) \rangle .
\label{chip}
\end{equation}
Bond structure factors and susceptibilities are defined in the same way;
we here consider those corresponding to correlations between bonds in the 
same lattice direction. Hence, defining $x_k$ and $y_k$ as the 
nearest-neighbor sites of site $k$ in the $x$- and $y$-directions, the
bond structure factors $S_{b,x}$ and $S_{b,y}$ are
\begin{equation}
S_{b,\alpha}(q_x,q_y) = {1\over N}\sum\limits_{k,l} 
{\rm e}^{i({\bf r}_k-{\bf r}_l)\cdot {\bf q}}
\langle B_{k,\alpha_k}B_{l,\alpha_l} \rangle, 
\label{sbaq}
\end{equation}
and clearly $S_{b,x}(q_x,q_y)=S_{b,y}(q_y,q_x)$. The corresponding 
susceptibilities are again defined as in Eq.~(\ref{chip}).

For expectation values involving products of operators that also appear as
terms in the Hamiltonian, such as the above plaquette and bond structure 
factors and susceptibilities, the SSE estimators are remarkably simple 
expressions involving only numbers of operators or operator combinations 
in the list $S_n$.\cite{sse2} The simplest case is the expectation value of 
a single operator,
\begin{equation}
\langle H_{t,a}\rangle = {\langle n([a,b])\rangle \over \beta},
\end{equation}
where $n([a,b])$ is the number of elements $[a,b]$ in the list $S_n$. This
gives the internal energy 
\begin{equation}
E = -{\langle n\rangle \over \beta},
\end{equation}
which is identical to the expression obtained by Handscomb.\cite{handscomb}
An equal-time correlation function of two operators appearing in the 
Hamiltonian is given by \cite{sse2}
\begin{equation}
\langle H_{s,a}H_{t,b}\rangle = {1\over \beta^2} \langle 
(n-1)N([s,a][t,b]) \rangle ,
\label{hhc}
\end{equation}
where $N([s,a][t,b])$ denotes the number of occurrences of the operators 
$[s,a]$ and $[t,b]$ next to each other, in the given order, in $S_n$ (with 
the periodicity of $S_n$ taken into account). The corresponding Kubo integral
is \cite{sse2}
\begin{eqnarray}
&&\int\limits_0^\beta d\tau \langle H_{s,a}(\tau)H_{t,b}(0)\rangle = \\ 
&&~~~{1\over \beta}\langle N([s,a])N([t,b]) - \delta_{st}\delta_{ab}
N([s,a]) \rangle, \nonumber
\label{hhx}
\end{eqnarray}
where $N([s,a])$ is the number of operators $[s,a]$. Using Eqs.~(\ref{hhc})
and (\ref{hhx}), the estimators for (\ref{strp})-(\ref{sbaq}) can be
easily obtained. 

\begin{figure}
\includegraphics[width=7cm]{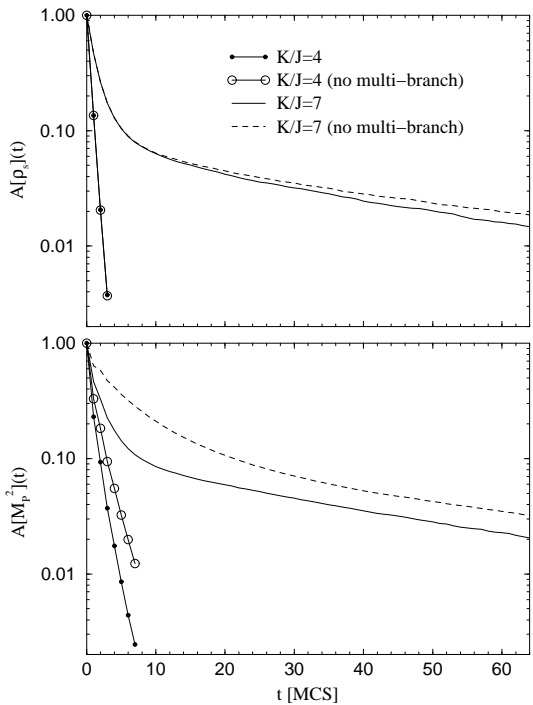}
\caption{Autocorrelation function for the spin stiffness (upper panel)
and the plaquette-stripe order parameter (lower panel) in simulations
of $L=16$ systems at $K/J=4$ and $7$, both at inverse temperature $K/T=16$.
Results of simulations both with and without the multi-branch cluster
update are shown.}
\label{fig:auto47}
\end{figure}

\section{Autocorrelations}
\label{AC}

We here show some results illustrating the performance of the algorithm,
focusing in particular on the efficiency boost achieved with the
multi-branch update. It would clearly be interesting to extract the dynamic 
exponent of the simulations at the various phase transitions, but we will not
attempt this here. Instead, we will focus on the simulation dynamics inside
the ordered phases. Particularly in the striped and staggered phases, which
break spatial symmetries, we expect slow modes corresponding to transitions 
between the different degenerate states. One might also expect potential 
problems related to long-lived defects forming in these states.

\begin{figure}
\includegraphics[width=7cm]{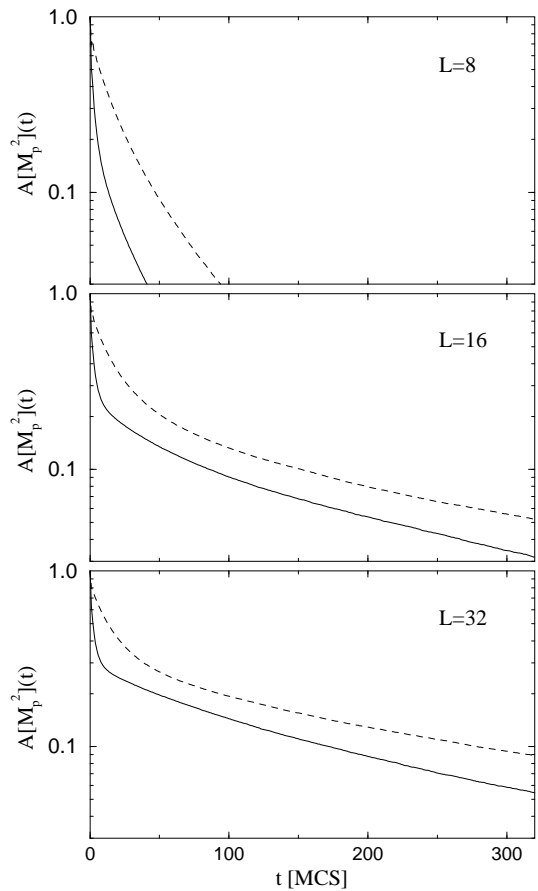}
\caption{Autocorrelation function for the plaquette-stripe order parameter
at $K/J=12$ and inverse temperature $K/T=32$. Results with (solid curves) 
and without (dashed curves) multi-branch clusters are compared for three
different system sizes.} 
\label{fig:auto12}
\end{figure}

For a quantity $Q$, the normalized autocorrelation function is defined in the
standard way as
\begin{equation}
A[Q](t) = \frac{\langle Q(i+t)Q(i)\rangle - \langle Q(i)\rangle ^2}
{\langle Q(i)^2\rangle}
\end{equation}
where the averages are over the Monte Carlo time (steps) $i$. We will compare 
autocorrelation functions in the three different ordered phases, obtained
in simulations with and without multi-branch cluster updates. A Monte Carlo
step is defined as a full sweep of diagonal updates, followed by a number 
of directed-loop updates, and, if multi-branch updates are carried out, 
decomposition of the configuration into clusters, each of which is flipped
with probability $1/2$. In these simulations the number of directed-loop
updates per step was chosen so that, on average, the total number of
vertices visited is $4M$, with the truncation $M$ of the index sequence
chosen equal to $1.25$ times the maximum expansion order $n$ reached during
equilibration (the dependence of $M$ on the length of the equilibration is
in practice very small and introduces only a negligible ambiguity in the
definition of the Monte Carlo time).

\begin{figure}
\includegraphics[width=7cm]{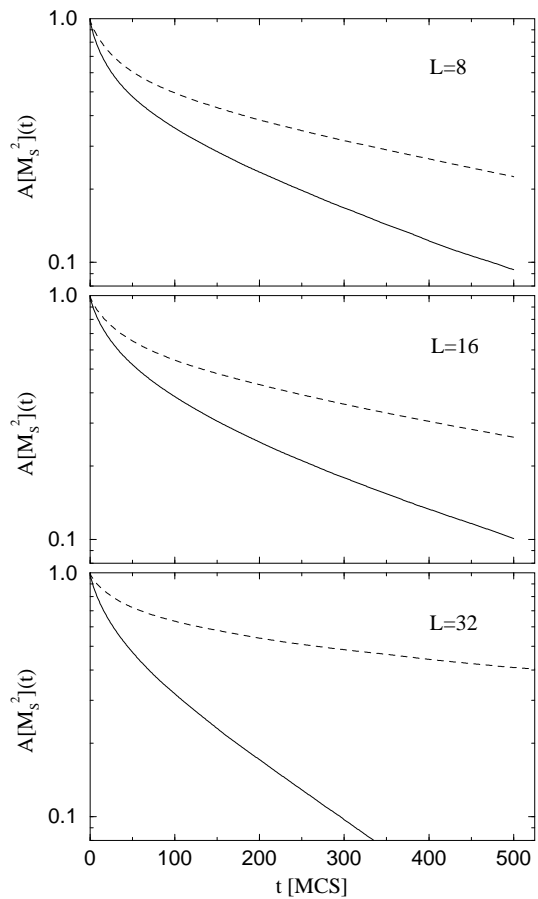}
\caption{Autocorrelation function for the staggered order parameter
at $K/J=32$ and inverse temperature $K/T=32$. Results with (solid curves) 
and without (dashed curves) multi-branch clusters are compared for three
different system sizes.}
\label{fig:auto32}
\end{figure}

Fig.~\ref{fig:auto47} shows autocorrelation results for the superfluid
density $\rho_s$ and the squared stripe-order-parameter $M_P^2$ inside the 
superfluid phase for a $16\times 16$ lattice at $K/T=16$. At $K/J=4$, the 
$\rho_s$ autocorrelations drop very rapidly (the integrated autocorrelation
time is less than $1$), and there are no discernible effects of including 
multi-branch updates. The autocorrelation time for $M_P^2$ is also very 
short, but here there are clear improvements with the multi-branch updates. 
However, considering that the CPU time is almost doubled when including 
multi-branch updates, including them at $K/J=4$ is not advantageous.
At $K/J=7$, which is approaching the transition point to the striped phase 
at $K/J \approx 7.9$, the autocorrelations decay much slower, and although 
there are visible favorable effects of the multi-branch updates in both 
quantities, the gain is hardly worth the additional CPU time cost. In 
Fig.~\ref{fig:auto12}, results are shown for the stripe order parameter 
at $K/J=12$, well inside the striped phase, for three different system 
sizes at inverse temperature $K/T=32$. The multi-branch updates have clear 
favorable effects on the autocorrelations, but although the initial drop is 
considerably faster, the asymptotic autocorrelation time, i.e., the long-time
linear decay seen on the linear-log scale used in the figures, changes 
very little. In this case the reduction of the integrated autocorrelation time
may (depending on the exact value of $K/J$, and the system size) motivate the 
additional computational effort of the multi-branch update. 

\begin{figure}
\includegraphics[width=7cm]{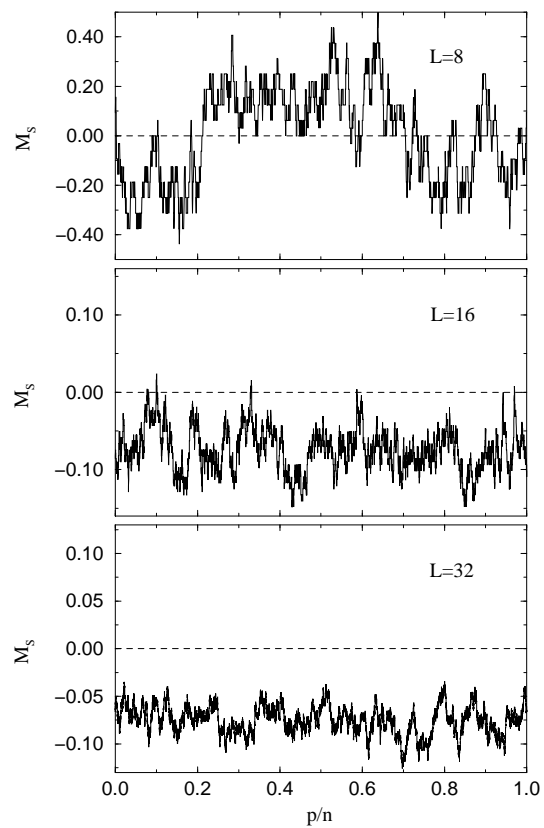}
\caption{The staggered magnetization vs the propagation index $p$ (divided
by the total number of operators $n$) in configurations generated for system
sizes $L=8,16$, and $32$, at $K/J=32$, $K/T=32$.}
\label{fig:mfluct}
\end{figure}

\begin{figure}
\includegraphics[width=7cm]{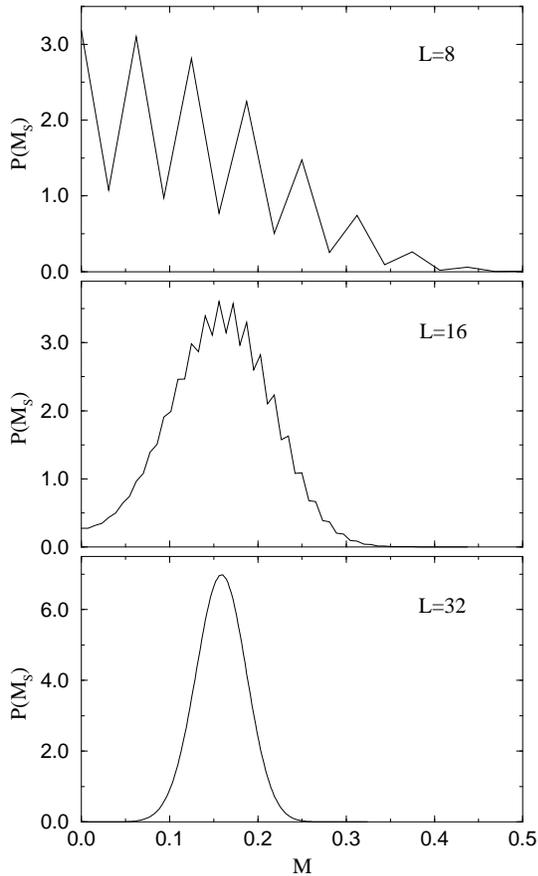}
\caption{Distribution of the staggered magnetization at $K/J=32$, $K/T=32$.}
\label{fig:mhist}
\end{figure}

The multi-branch cluster update improves the simulation efficiency
considerably inside the CDW phase, as illustrated in Fig.~\ref{fig:auto32}
for three different system sizes with $K/J=32$ at a low temperature.
Here the improvement in the simulation efficiency for the squared staggered
order parameter $M_S^2$ is clearly significant enough to motivate the cost
of the multi-branch clusters, especially for large system sizes. An 
interesting feature to note here is that when the multi-branch clusters
are included, the asymptotic autocorrelation time actually decreases for
$L=32$ relative to $L=16$, and $L=16$ and $L=8$ show almost identical
autocorrelation functions. This surprising trend for increasing $L$ can
probably be traced to the fluctuations in the CDW order parameter for
a given SSE configuration. Fig.~\ref{fig:mfluct} shows the dependence
of the staggered order parameter on the propagation number $p$ [referring
to the propagated states, Eq.~(\ref{propagated})] divided by the total 
number of operators $n$ for an equilibrated configuration. The fraction 
$p/n$ corresponds roughly to the normalized imaginary-time $\tau/\beta$ 
in the standard Euclidean path integral formalism.\cite{irsse} For a
small system, exemplified by $L=8$ in the figure, the order parameter
fluctuates between positive and negative values, whereas for a large
system, exemplified here by $L=32$, fluctuations sufficiently large
to ``tunnel'' the system between positive and negative order parameters
are very rare. Clearly, as $T\to 0$, there would be such tunneling
events also in a large system, but if $T$ is not low compared to the
gap between the symmetric and antisymmetric linear combinations of
the two different real-space ordered states (which decreases exponentially
fast with increasing $L$), such events are not present in typical
configurations. The shorter autocorrelation time for $L=32$ than for
$L=16$ (when multi-branch updates are included) in Fig.~\ref{fig:auto32}
may hence be related to the larger fluctuations in $M_S$ for the
smaller system size, which can lead to various tunneling events that
are not so easily added or removed from the configurations. For the
larger system size, there are in practice no tunneling events at the
temperature used here, and the difficulties in adding/removing them
in this case would only show up at very long times as an undetectable
tail in the autocorrelation function. The order parameter fluctuations
are further illustrated in the form of histograms in Fig.~\ref{fig:mhist}
(only showing the positive part of the distributions). Here it can be
seen that some tunneling events are still expected at $L=16$, since
the probability of a zero order parameter is not negligible, but at
$L=32$ the probability at zero is exponentially small and hardly any
tunneling events would be expected on the time scale of a typical
simulation.

\begin{figure}
\includegraphics[width=7cm]{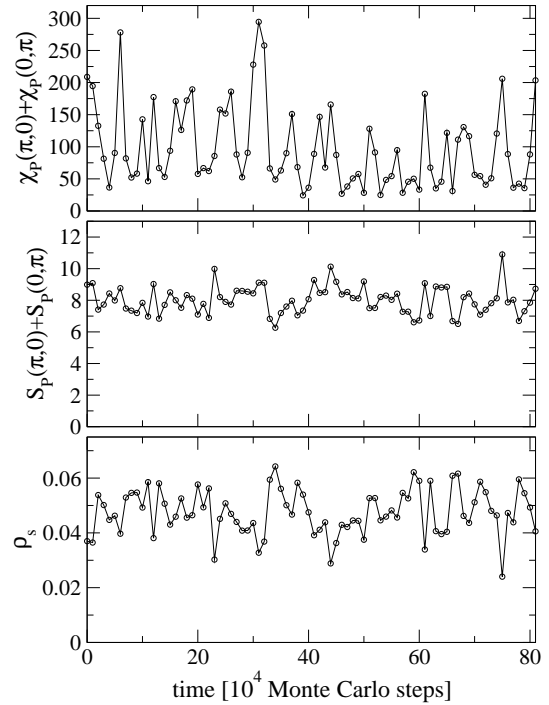}
\caption{Bin averages for the superfluid stiffness, the stripe structure
factor, and the stripe susceptibility, for an $L=96$ lattice at $K/J=7.91$
and inverse temperature $K/T=64$. Each point represents an average over
$10^4$ Monte Carlo steps.}
\label{fig:mctime}
\end{figure}

As can be seen in Figs.~\ref{fig:auto47}, the asymptotic autocorrelation 
time for the stripe order is quite long, approximately 40 Monte Carlo steps,
in the superfluid phase at $K/J=7$ even for the modest system size $L=16$. 
Critical slowing down is expected as the critical superfluid-striped point
is approached. Although we have not yet attempted to extract the 
corresponding dynamic critical exponent, we show, in Fig.~\ref{fig:mctime},
results for the Monte Carlo time evolution of some relevant critical 
quantities very close to the quantum phase transition. Here the
lattice size $L=96$ and the temperature is chosen sufficiently low for 
obtaining ground state expectation values within statistical error. 
The points shown are averages over ``bins'' of $10^4$ Monte Carlo steps, and 
clearly these bin averages are not yet statistically independent; the 
autocorrelation times are several $10^4$ Monte Carlo steps. These simulations
did include multi-branch cluster updates. Currently, high-precision 
$T\to 0$ converged simulations of this model close to the superfluid-VBS
transition are not feasible for $L$ much larger than $100$.

Note the clear anti-correlations between the stripe order and superfluid
density in Fig.~\ref{fig:mctime}. These do not, however, give an indication
of the order of the phase transition between the two phases, as anti-correlations
are expected at both continuous and first-order transitions.

\section{Diagonal plaquette interactions}

In general, diagonal terms can be added to the Hamiltonian Eq.~(\ref{ham}) 
without the development of a new directed-loop algorithm; only the exit
probability tables change, due to the modified weights of the relevant
diagonal (C) vertices illustrated in Fig.~\ref{fig:xyvert} (and symmetry 
related sets).  We here consider three different diagonal terms: (i) one 
which enhances or suppresses staggered  (``flippable'' by the K-term) 
plaquettes, (ii) a uniform external magnetic (Zeeman) field, and
(iii) a staggered field.

\subsection{Flippable-plaquette interaction}

The full spin Hamiltonian including the flippable-plaquette interaction is 
\begin{equation}
H = -J \sum_{\left< ij \right>}B_{ij} - K \sum_{\left< ijkl \right>}P_{ijkl}
 \\  -  V\sum_{\left< ijkl \right>} Q_{ijkl},
\label{hdiagV}
\end{equation}  
where the bond ($B_{ij}$) and plaquette ($P_{ijkl}$) operators are defined 
in Eqs.~(\ref{bond}) and (\ref{plaquette}) and $Q_{ijkl}$ is $1$ or $0$
for flippable and non-flippable plaquettes, respectively, or  
\begin{eqnarray}
Q_{ijkl} &=& \left({{1}/{2}+S^z_i}\right) \left({{1/}{2}-S^z_j}\right) 
\left({{1}/{2}+S^z_k}\right) \left({{1}/{2}-S^z_l}\right) \nonumber \\
&+& \left({{1}/{2}-S^z_i}\right) \left({{1}/{2}+S^z_j}\right) 
\left({{1}/{2}-S^z_k}\right) \left({{1}/{2}+S^z_l}\right). \nonumber
\label{Qijkl}
\end{eqnarray}
This $V$ interaction is interesting as it produces an exactly soluble 
(Rokhsar-Kivelson\cite{RK}) point at $J=0$ and  $-K=V$,
and is similar to the term employed in quantum dimer models\cite{RSrk} 
(see also Ref.~\onlinecite{balents}). Hence, its usefulness making a connection 
between numerical and analytical studies of microscopic Hamiltonians is 
immediately obvious.

To solve the directed loop equations in the presence of a diagonal interaction
such as the $V$ term, the general procedure is simply to identify, and 
modify, the relevant sets of directed-loop equations which include the vertex 
weighted by $V$.  Here, the diagonal term $H_{1,a}$ of Eq.~(\ref{hab})
becomes $H_{1,a} =  C I_{ijkl}+V Q_{ijkl}$. In the SSE formalism, these 
diagonal matrix elements are represented as C-vertices, each vertex having 
a weight given by Eq.~(\ref{wp}). The modified vertices of interest here
are denoted
\begin{eqnarray}
\label{WV}
W^V &=& \langle{\da\ua\da\ua \left|{H_a}\right| \da\ua\da\ua}\rangle  
= \langle{\ua\da\ua\da \left|{H_a}\right| \ua\da\ua\da}\rangle  \\ \nonumber
&=& V + C ,
\end{eqnarray}
where we have represented the plaquettes of the basis state 
$| \alpha \rangle$ by a list of 
the spin states in the order $ijkl$ corresponding to Fig.~\ref{fig:labels}(a).
The directed-loop equations that are relevant to this diagonal term are 
related to those illustrated in Fig.~\ref{fig:xyvert}, but clearly only the 
closed sets that contain fully-staggered vertices are affected by $V$.
The set shown in Fig.~\ref{fig:xyvert}
does not have any such vertices and hence the corresponding 
directed-loop solution is the same as with $V=0$. In Fig.~\ref{VSv} we show
a closed set that is affected by $V$. Here we have included the bounce
processes because they can no longer be completely excluded.  The 
directed-loop equations for this set are written as
\begin{eqnarray}
\label{hdirloopV}
v_{11}+v_{12}+v_{13}+v_{14} &=& W_1 = C \nonumber, \\ 
v_{21}+v_{22}+v_{23}+v_{24} &=& W^V = V + C \nonumber, \\ 
v_{31}+v_{32}+v_{33}+v_{34} &=& W_3 = {J}/{2} , \\ 
v_{41}+v_{42}+v_{43}+v_{44} &=& W_4 = {J}/{2} , \nonumber 
\end{eqnarray}
where $W_1$, $W_3$ and $W_4$ are the same as those given before in 
Eq.~(\ref{eqset1}). In order to solve these equations, we use the same 
detailed balance requirements, Eq.~(\ref{balancereq}), and symmetry 
arguments as previously to constrain the equations and produce a unique 
solution. Our choice of symmetry conditions here correspond to 
\begin{eqnarray}
\label{symmv}
v_{12}&=&v_{13}, \nonumber \\ 
v_{22}&=& v_{23}, \\ 
v_{34} &=& v_{44}. \nonumber
\end{eqnarray}
Two forms of the solutions are needed in order to ensure positive-definite 
vertex weights for all choices of parameters. The first solution is valid 
for small couplings, $|V| \leq J$, and can be formulated without the 
undesirable bounce processes (the right-hand column):
\begin{widetext}
\begin{equation}
\begin{array}{llll}
v_{11} = C-J/2+V/2, &v_{12} = J/4 -V/4, &v_{13} = J/4 -V/4, &v_{14} = 0,\\
v_{21} = C-J/2+V/2, &v_{22} = J/4 +V/4, &v_{23} = J/4 +V/4, &v_{24} = 0,\\
v_{31} = J/4 -V/4, &v_{32} = J/4 +V/4, &a_{33} = 0, &v_{34} = 0, \\
v_{41} = J/4 -V/4, &v_{42} = J/4 +V/4, &a_{43} = 0, &v_{44} = 0.
\end{array}
\label{ESv1}
\end{equation}
where we see that the constant $C$ must be greater than $J/2-V/2$ to ensure 
that $v_{11}$ remains positive.   However, in order to satisfy the requirement
that $a_{11}$ is positive for vertex-path sets not affected by the $V$ term,
one should set $C>J/2$ for the case $V>0$.  For $V<0$, one must in addition 
ensure that the weight $W^V$ in Eq.~(\ref{hdirloopV}) remains positive, 
requiring a constant $C>J/2+|V|/2$.
Note that in the limit $V \rightarrow 0$, this
equation set is equal to the directed-loop equations previously obtained
in Eq.~(\ref{cjsolution}). 
Clearly, when $|V| > J$, a different form 
of the solutions must be constructed in order to ensure that each vertex 
weight stays positive. In the case $V>J>0$, a non-zero bounce probability 
$v_{24} = V - J$ must be included.  A convenient form of the solution in 
this case is then
\begin{equation}
\begin{array}{llll}
v_{11} = C, &v_{12} = 0, &v_{13} = 0, &v_{14} = 0,  \\
v_{21} = C, &v_{22} = J/2, &v_{23} = J/2, &v_{24} = V-J, \\
v_{31} = 0, &v_{32} = J/2, &v_{33} = 0, &v_{34} = 0, \\
v_{41} = 0, &v_{42} = J/2, &v_{43} = 0, &v_{44} = 0,
\end{array}
\label{ESv3}
\end{equation}
where the bounce process is turned on slowly, i.e.~linearly with $V-J$, 
ensuring a small bounce probability in the algorithm for a moderate range 
of $V$ larger than the exchange.  
However, it is clear that the bounce 
process $v_{24}$ becomes negative for negative $V$ (i.e. $V<-J<0$) and we 
hence need a different solution in this case. Again, we use a solution 
that turns the bounce processes on slowly, but in this case two non-zero 
bounces are required:
\begin{equation}
\begin{array}{llll}
v_{11} = C+J-2|V|, &v_{12} = J/2, &v_{13} = J/2, &v_{14} = 2|V|-2J,  \\
v_{21} = C+J-2|V|, &v_{22} = 0, &v_{23} = 0, &v_{24} = |V|-J, \\
v_{31} = J/2, &v_{32} = 0, &v_{33} = 0, &v_{34} = 0, \\
v_{41} = J/2, &v_{42} = 0, &v_{43} = 0, &v_{44} = 0.
\end{array}
\label{ESv4}
\end{equation}
\begin{center}
\begin{figure}
\includegraphics[height=4cm]{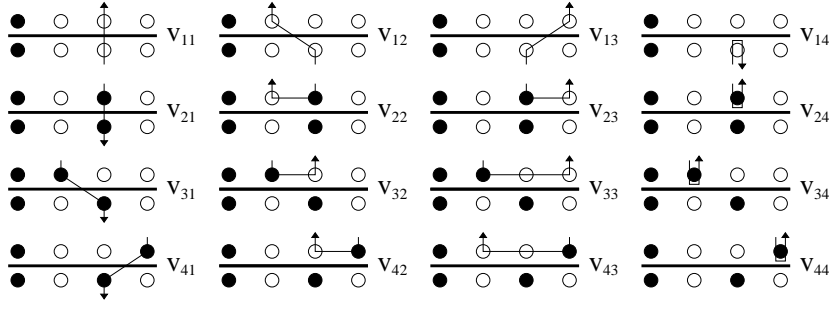}
\caption{A closed set of C and J vertex-paths for which the corresponding
directed-loop equations are different for $V=0$ and $V\not=0$. A second set 
of vertex-paths---the spin-reverse of the one shown here---also gives the 
same directed-loop solution.}
\label{VSv}
\end{figure}
\end{center}
\end{widetext}
This last equation set imposes the requirement that $C > 2|V|-J$.
In Eqs.~(\ref{ESv1}) to (\ref{ESv4}), the actual exit probabilities for 
the directed-loop algorithm are obtained in the usual way by dividing the 
matrix elements by the vertex weights; $P^v_{ij} = v_{ij}/W_i$, 
where $W_{i}$ is the relevant matrix element. 

Note again that when implementing a diagonal interaction such as the $V$
term,
the only change required in the simulation code,
relative to the pure J-K model, 
is the probability weights of only the specific relevant vertex-paths
affected.
In the case above for the plaquette $V$ term, only the vertex set 
shown in Fig.~\ref{VSv}, and the related set with the other staggered 
vertex, will use the solutions outlined in Eqs.~(\ref{ESv1}), (\ref{ESv3})
and (\ref{ESv4}).  All other C to J vertex sets which do not contain 
fully-staggered diagonal vertices will use the original solution, 
Eq.~(\ref{cjsolution}).

\subsection{Uniform magnetic field}

Perhaps the simplest extension of the J-K Hamiltonian (\ref{ham}) is the 
addition of an external magnetic field.  A diagonal Zeeman field $h$ 
coupling to the $z$-components of the $S=1/2$ spins is of particular 
physical importance, as it cants the magnetization away from the 
zero-magnetization state, or dopes the system away from half-filling in the 
boson language.\cite{melko}
The possibility of deconfined quantum critical points occurring between 
superfluid and insulating states at commensurate fillings other than 
one-half is also currently a question of interest.\cite{leonandanton}  The 
Hamiltonian under study is now
\begin{equation}
H = -J \sum_{\left< ij \right>}B_{ij} - K \sum_{\left< ijkl
\right>}P_{ijkl} - h \sum_i S_i^z ,
\label{hunifH}
\end{equation}  
where we restrict $h>0$.
The diagonal term $H_{1,a}$ of Eq.~(\ref{hab}) is modified to include 
the effects of the field,
\begin{equation}
H_{1,a} = \frac{h}{4} \left({ S_i^z+S_j^z+S_k^z+S_l^z }\right) 
+ C I_{ijkl},
\label{hH1a}
\end{equation}
where this term now produces different matrix elements depending on 
the spins $S_i^z,S_j^z,S_k^z,S_l^z$, with an associated vertex 
weight, Eq.~(\ref{wp}).  We can ensure that each weight will remain 
positive by adjusting $C$, in particular, we write $C=h/2 + \epsilon$, 
where $\epsilon >0$ is typically a small constant.  Representing the relevant
plaquette of the state $| \alpha \rangle$ by a list of the spin states, 
we can calculate the weights of the 16 C-vertices using Eq.~(\ref{hH1a}).  
The results are summarized in Table~\ref{Uweights}.

\begin{table}
\caption{The weight factors for the diagonal vertices in the uniform J-K model.}
\label{Uweights}
\begin{ruledtabular}
\begin{tabular}{cp{2mm}cp{2mm}c}
  && $\langle{ S_i^zS_j^zS_k^zS_l^z \left|{H_a}\right| S_i^zS_j^zS_k^zS_l^z}\rangle$
&& weight factor \\ \hline \\
$W^h_1$ && $\langle{\ua\ua\ua\ua \left|{H_a}\right| \ua\ua\ua\ua}\rangle$ && $h +  \epsilon$ \\
$W^h_2$ && $\langle{\da\ua\ua\ua \left|{H_a}\right| \da\ua\ua\ua}\rangle$ && $3h/4 +  \epsilon$\\
$W^h_2$ && $\langle{\ua\da\ua\ua \left|{H_a}\right| \ua\da\ua\ua}\rangle$ && $3h/4 +  \epsilon$\\
$W^h_2$  && $\langle{\ua\ua\da\ua \left|{H_a}\right| \ua\ua\da\ua}\rangle$ && $3h/4 +  \epsilon$ \\
$W^h_2$  && $\langle{\ua\ua\ua\da \left|{H_a}\right| \ua\ua\ua\da}\rangle$ && $3h/4 +  \epsilon$ \\
$W^h_3$ && $\langle{\ua\ua\da\da \left|{H_a}\right| \ua\ua\da\da}\rangle$ && $h/2 +  \epsilon$ \\
$W^h_3$ && $\langle{\ua\da\da\ua \left|{H_a}\right| \ua\da\da\ua}\rangle$ && $h/2 +  \epsilon$ \\
$W^h_3$ && $\langle{\da\da\ua\ua \left|{H_a}\right| \da\da\ua\ua}\rangle$ && $h/2 +  \epsilon$ \\
$W^h_3$ && $\langle{\da\ua\da\ua \left|{H_a}\right| \da\ua\da\ua}\rangle$ && $h/2 +  \epsilon$ \\
$W^h_3$ && $\langle{\ua\da\ua\da \left|{H_a}\right| \ua\da\ua\da}\rangle$ && $h/2 +  \epsilon$ \\
$W^h_3$ && $\langle{\da\ua\ua\da \left|{H_a}\right| \da\ua\ua\da}\rangle$ && $h/2 +  \epsilon$ \\
$W^h_4$ && $\langle{\ua\da\da\da \left|{H_a}\right| \ua\da\da\da}\rangle$ && $h/4+\epsilon$ \\
$W^h_4$ && $ \langle{\da\ua\da\da \left|{H_a}\right| \da\ua\da\da}\rangle$ && $h/4+\epsilon$ \\
$W^h_4$ && $ \langle{\da\da\ua\da \left|{H_a}\right| \da\da\ua\da}\rangle $ && $h/4+\epsilon$ \\
$W^h_4$ && $ \langle{\da\da\da\ua \left|{H_a}\right| \da\da\da\ua}\rangle$ && $h/4+\epsilon$ \\
$W^h_5$ && $\langle{\da\da\da\da \left|{H_a}\right| \da\da\da\da}\rangle$ && $\epsilon$
\end{tabular} \end{ruledtabular}
\end{table}

In addition, with the Hamiltonian Eq.~(\ref{hunifH}), the off-diagonal plaquette
operators $H_{2,a}$ to $H_{6,a}$ in Eq.~(\ref{hab}) remain unmodified.  In this 
case, there are now four unique sets of directed loop equations for C$\to$C and 
C$\to$J vertices that are not related by trivial symmetry operations.  

The first closed set of C and J vertex paths is illustrated in 
Fig.~\ref{fig:xyvert}, with weights $a_{ij}$ which now also should
include the bounce processes $a_{i4}$ left out in the figure. If we 
recall that the open circle in Fig.~\ref{fig:xyvert} denotes a spin 
down, or $S^z = -1/2$, then, the directed-loop equations corresponding 
to this set is now modified from Eq.~(\ref{eqset1}) to read
\begin{eqnarray}
a_{11}+a_{12}+a_{13}+a_{14} &=& W^h_5 = \epsilon \nonumber, \\ 
a_{21}+a_{22}+a_{23}+a_{24} &=& W^h_4 = h/4+\epsilon \nonumber, \\ 
a_{31}+a_{32}+a_{33}+a_{34} &=& W_3 = {J}/{2} , \label{hdirloopA} \\ 
a_{41}+a_{42}+a_{43}+a_{44} &=& W_4 = {J}/{2}. \nonumber 
\end{eqnarray}
We use the same detailed balance, Eq.~(\ref{balancereq}) and symmetry arguments, 
Eq.~(\ref{symmv}), as previously to constrain the 
equation set and produce a unique solution.  For fields $h<4J$, we can 
obtain a solution which contains no bounce processes, 
\begin{widetext}
\begin{equation}
\begin{array}{llll}
a_{11} = h/8-J/2+\epsilon, &a_{12} = J/4 -h/16, &a_{13} = J/4 -h/16, &a_{14} = 0,
\\
a_{21} = h/8-J/2+\epsilon, &a_{22} = J/4 +h/16, &a_{23} = J/4 +h/16, &a_{24} = 0,
\\
a_{31} = J/4 -h/16, &a_{32} = J/4 +h/16, &a_{33} = 0, &a_{34} = 0, \\
a_{41} = J/4 -h/16, &a_{42} = J/4 +h/16, &a_{43} = 0, &a_{44} = 0,
\end{array}
\label{ESa1}
\end{equation}
\end{widetext}
where, to keep the element $a_{11}$ positive-definite, we require 
$\epsilon \geq J/2-h/8$.  
Again, for $h>4J$, the form of the solutions 
must change in a non-trivial way in order to keep each vertex weight 
positive.  This requirement produces a non-zero bounce process, $a_{24}$, 
which together with the other weights gives the high-field solution
%\begin{widetext}
\begin{equation}
\begin{array}{llll}
a_{11} = \epsilon, &a_{12} = 0, &a_{13} = 0, &a_{14} = 0,  \\
a_{21} = \epsilon, &a_{22} = J/2, &a_{23} = J/2, &a_{24} = h/4-J, \\
a_{31} = 0, &a_{32} = J/2, &a_{33} = 0, &a_{34} = 0, \\
a_{41} = 0, &a_{42} = J/2, &a_{43} = 0, &a_{44} = 0.
\end{array}
\label{ESa3}
\end{equation}
%\end{widetext}

The second independent closed set of vertex weights for the Hamiltonian 
Eq.~(\ref{hunifH}) is obtained by taking the spin-reverse of the closed 
set illustrated in Fig.~\ref{fig:xyvert}. The resulting directed-loop 
equations then contain the fully polarized vertex in Table~\ref{Uweights}, 
and are written as
\begin{eqnarray}
c_{11}+c_{12}+c_{13}+c_{14} &=& W^h_1 = h+\epsilon, \nonumber \\
c_{21}+c_{22}+c_{23}+c_{24} &=& W^h_2 = 3h/4+\epsilon, \nonumber \\
c_{31}+c_{32}+c_{33}+c_{34} &=& W_3 = {J}/{2} , \\
c_{41}+c_{42}+c_{43}+c_{44} &=& W_4 = {J}/{2} . \nonumber
\end{eqnarray}
This set is solved in the same way as set $a$ above, employing analogous 
conditions for detailed-balance and vertex symmetries.  The result is two 
sets of vertex weights; the first, for $h<4J$ is
\begin{widetext}
\begin{equation}
\begin{array}{llll}
c_{11} = 7h/8-J/2+\epsilon, &c_{12} = J/4 +h/16, &c_{13} = J/4 +h/16, &c_{14} = 0,
\\
c_{21} = 7h/8-J/2+\epsilon, &c_{22} = J/4 -h/16, &c_{23} = J/4 -h/16, &c_{24} = 0,
\\
c_{31} = J/4 +h/16, &c_{32} = J/4 -h/16, &c_{33} = 0, &c_{34} = 0, \\
c_{41} = J/4 +h/16, &c_{42} = J/4 -h/16, &c_{43} = 0, &c_{44} = 0,
\end{array}
\label{ESb1}
\end{equation}

\begin{center}
\begin{figure}
\includegraphics[height=4cm]{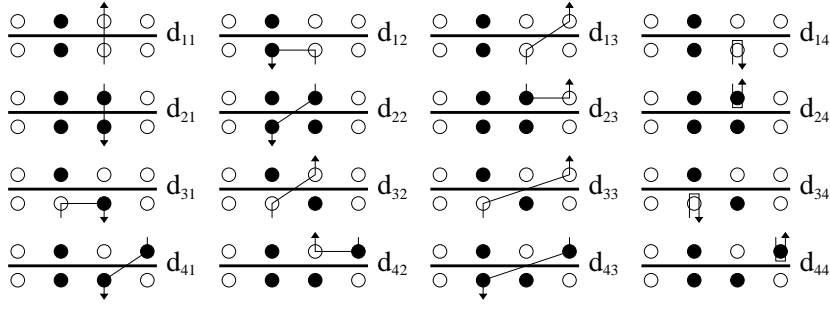}
\caption{A closed set of C and J vertex-paths used in solving the directed-loop equations for the uniform-field J-K-$h$ model.}
\label{VSd}
\end{figure}
\end{center}

\end{widetext}
where clearly, to keep $c_{11}>0$ we require $\epsilon \ge J/2-7h/8$.  
We see that this solution is similar in form to the low-field solution 
for vertex set $a$, Eq.~(\ref{ESa1}).  However, the  $c_{11}$ and $c_{21}$ 
terms are modified, and also the other non-zero terms have the opposite 
sign of $h$ (i.e. if $a_{ij} = J/4 \pm h/16$, then $c_{ij} = J/4 \mp h/16$).
The other solution corresponds to the large-field case, $h>4J$, and requires 
the inclusion of a non-zero weight for the bounce process $c_{14}$,
\begin{equation}
\begin{array}{llll}
c_{11} = \epsilon+3h/4, &c_{12} = J/2, &c_{13} = J/2, &c_{14} = h/4-J,  \\
c_{21} = \epsilon+3h/4, &c_{22} = 0, &c_{23} = 0, &c_{24} = 0, \\
c_{31} = J/2, &c_{32} = 0, &c_{33} = 0, &c_{34} = 0, \\
c_{41} = J/2, &c_{42} = 0, &c_{43} = 0, &c_{44} = 0.
\end{array}
\label{ESb3}
\end{equation}
Again, the $c_{11}$ and $c_{21}$ terms are modified by the different 
vertex sets, and in addition the non-zero bounce process has moved to 
$c_{14}$ from $a_{24}$ in Eq.~(\ref{ESa3}).

The third independent closed set of vertex weights does not include 
the fully spin-up or spin-down matrix elements of the previous two, 
sets $a$ and $c$.  The diagrammatic representation is therefore not 
trivially related to these previous cases, and is illustrated in 
Fig.~\ref{VSd}.  In this case, the directed-loop equations are
\begin{eqnarray}
d_{11}+d_{12}+d_{13}+d_{14} &=& W^h_4 = h/4+\epsilon, \nonumber \\
d_{21}+d_{22}+d_{23}+d_{24} &=& W^h_3 = h/2+\epsilon, \nonumber \\
d_{31}+d_{32}+d_{33}+d_{34} &=& W_3 = {J}/{2},  \\
d_{41}+d_{42}+d_{43}+d_{44} &=& W_4 = {J}/{2}. \nonumber
\end{eqnarray}
While the detailed balance conditions for this equation set is the same as 
before, Eq.~(\ref{balancereq}), it can be noted that the additional 
symmetry conditions, Eq.~(\ref{symmv}), do not appear in the diagrams
here.  Although not immediately justifiable in terms of symmetry arguments, 
there is in general no reason why the same constrains cannot be used to solve 
equation set $d$, and therefore we will continue to use Eq.~(\ref{symmv}) as 
it facilitates implementation of the algorithm (although there is no
guarantee that this leads to the most efficient simulation).  The low-field 
solution ($h<4J$) is then given by the same equation set as solution 
$a$, Eq.~(\ref{ESa1}) with all $d_{ij}=a_{ij}$ {\it except} the following:
\begin{equation}
d_{11} = d_{21}=3h/8-J/2+\epsilon.
\label{ESc1}
\end{equation}
Here, it is quite obvious that we require $\epsilon \ge J/2-3h/8$ in order 
to keep all vertex weights positive.  For the high-field case ($h>4J$), we are 
forced to have a non-zero bounce process, and upon solving we again get an 
equation set similar to solution $a$, Eq.~(\ref{ESa3}), with the 
exception that
\begin{equation}
d_{11} =d_{12}= \epsilon+h/4.
\label{ESc3}
\end{equation}

The final set of vertex-weights used in the uniform-field solution is obtained 
by taking the spin-reverse of the closed set illustrated in Fig.~\ref{VSd}, in 
an analogous manner to the way that set $c$ was obtained from set $a$.  The 
result is the directed loop equations given by
\begin{eqnarray}
e_{11}+e_{12}+e_{13}+e_{14} &=& W^h_2 = 3h/4+\epsilon \nonumber, \\
e_{21}+e_{22}+e_{23}+e_{24} &=& W^h_3 = h/2+\epsilon \nonumber, \\
e_{31}+e_{32}+e_{33}+e_{34} &=& W_3 = {J}/{2},  \\
e_{41}+e_{42}+e_{43}+e_{44} &=& W_4 = {J}/{2}. \nonumber
\end{eqnarray}  
Again, we employ the detailed balance and symmetry conditions discussed above 
to find a unique low-field solution for $h<4J$, which in this case is the same 
as the solution set $c$, Eq.~(\ref{ESb1}) with the exception that
\begin{equation}
e_{11} =e_{21}= 5h/8-J/2+\epsilon,
\label{ESd1}
\end{equation}
where again the bounces have been eliminated, and to get $e_{11}>0$, we need 
$\epsilon \ge J/2-5h/8$.  The large-field ($h>4J$) solutions are given by the 
analogous set, Eq.~(\ref{ESb3}), with the exception that
\begin{equation}
e_{11}=e_{21} = \epsilon+h/2.
\label{ESd3}
\end{equation}

As before, the above four equations sets, $a$, $c$, $d$ and $e$, serve to 
uniquely define the exit probabilities,
given by dividing the matrix elements by the vertex weights, e.g.
$P^a_{ij} = a_{ij}/W_i$, where the values of $W_{i}$ are given either
by Table~\ref{Uweights} above, or by the previously-defined values
of $W_3 = W_4 = J/2$.
It is important, in the implementation of 
the directed-loop equations, that all diagonal vertices are weighted 
according to their proper equation set.  The relation of a general C 
or J vertex to the proper equation set in some cases depends on the 
path that a loop segments takes through the vertex. For example, the 
vertex in $a_{31}$ is related to $d_{31}$, however the path taken by 
the loop in each case results in a C vertex which is weighed differently 
by the field.  Also, we note that the common element of all of these equation 
sets is the constant $\epsilon$.  This $\epsilon$ must be chosen to keep 
all of the elements $a_{11}, c_{11}, d_{11}, e_{11}$ and their symmetry 
related weights positive definite.  The critical condition comes from 
Eq.~(\ref{ESa1}), where $a_{11}>0$ in all cases for $\epsilon \ge J/2-h/8$.  
It can be seen that, if this condition is satisfied, then all of the weights 
$c_{11}, d_{11}$ and $e_{11}$ will automatically be positive definite, and 
it is therefore the $\epsilon$ that we choose in implementation of the algorithm.

\subsection{Staggered magnetic field}

The final set of directed loop solutions that we will present in this paper 
is for the J-K model in a staggered Zeeman field.  Motivation for this 
extension of the Hamiltonian comes directly from predictions in the theory 
of deconfined quantum criticality \cite{senthil,ashvin} and its applicability 
to our microscopic model.  In short, a staggered Zeeman field on our spin 
model corresponds to a uniform Zeeman field that couples to the $z$ component 
of $\hat{n}$ in the nonlinear-sigma model of relevance.  The theory then 
predicts a ``split'' transition between the VBS and superfluid phases, with 
an intermediate phase with neither order (but with a ``background'' field-induced
staggered magnetization).
\begin{center}
\begin{figure}
\includegraphics[height=4cm]{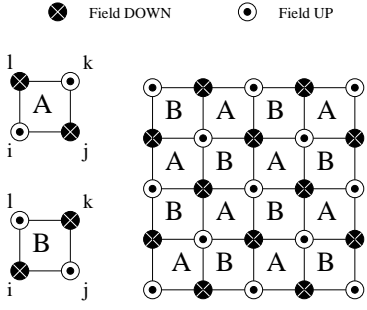}
\caption{Sublattice decoration of the two-dimensions square lattice, used in constructing the quantum Monte Carlo simulation of the J-K model with staggered Zeeman field. }
\label{SL}
\end{figure}
\end{center}

The modified Hamiltonian is
\begin{equation}
H = -J \sum_{\left< ij \right>}B_{ij} - K \sum_{\left< ijkl
\right>}P_{ijkl} - h \sum_i  (-1)^{x_i+y_i} S_i^z ,
\label{hstaggH}
\end{equation}
where $x_i$ and $y_i$ are the Cartesian lattice coordinates of the $i$th spin, 
and $h>0$.
The diagonal plaquette term $H_{1,a}$ of Eq.~(\ref{hab}) is modified to include 
the effects of the staggered field,
\begin{equation}
H_{1,a} = (-1)^{x_i+y_i} \frac{h}{4} \left({ S_i^z-S_j^z+S_k^z-S_l^z }\right) +
C I_{ijkl},
\label{hSt1}
\end{equation}
and the other plaquette Hamiltonian terms remain unmodified.  Keeping $C$ 
arbitrary for now, we can easily calculate the weights for the 16 diagonal 
(C) vertices. The approach we take in constructing the simulation is the 
decorate the lattice with an ``A'' and ``B'' sublattice in a checkerboard pattern
(Fig.~\ref{SL}).  The solution to each vertex weight in the directed loop
equations will have two components, one if the vertex happens to fall on
an ``A'' plaquette, and another for the same vertex on a ``B'' plaquette
(see Table~\ref{ABweights}).

\begin{table}
\caption{The weight factors for the diagonal vertices in the staggered J-K model.}
\label{ABweights}
\begin{ruledtabular}
\begin{tabular}{cp{2mm}cp{2mm}c}
$\langle{ S_i^zS_j^zS_k^zS_l^z \left|{H_a}\right| S_i^zS_j^zS_k^zS_l^z}\rangle$
&& A sublattice && B sublattice\\ \hline \\
$\langle{\ua\ua\ua\ua \left|{H_a}\right| \ua\ua\ua\ua}\rangle$
&&  $C$ &&  $C$ \\
$\langle{\da\da\da\da \left|{H_a}\right| \da\da\da\da}\rangle$
&&  $C$ &&  $C$ \\
$\langle{\da\ua\ua\ua \left|{H_a}\right| \da\ua\ua\ua}\rangle$
&& $-h/4+C$ && $h/4+C$ \\
$\langle{\ua\da\ua\ua \left|{H_a}\right| \ua\da\ua\ua}\rangle$
&& $h/4+C$ && $-h/4+C$ \\
$\langle{\ua\ua\da\ua \left|{H_a}\right| \ua\ua\da\ua}\rangle$
&& $-h/4+C$ && $h/4+C$ \\
$\langle{\ua\ua\ua\da \left|{H_a}\right| \ua\ua\ua\da}\rangle$
&& $h/4+C$ && $-h/4+C$ \\
$\langle{\ua\da\da\da \left|{H_a}\right| \ua\da\da\da}\rangle$
&& $h/4+C$ && $-h/4+C$ \\
$\langle{\da\ua\da\da \left|{H_a}\right| \da\ua\da\da}\rangle$
&& $-h/4+C$ && $h/4+C$ \\
$\langle{\da\da\ua\da \left|{H_a}\right| \da\da\ua\da}\rangle$
&& $h/4+C$ && $-h/4+C$ \\
$\langle{\da\da\da\ua \left|{H_a}\right| \da\da\da\ua}\rangle$
&& $-h/4+C$ && $h/4+C$ \\
$\langle{\ua\ua\da\da \left|{H_a}\right| \ua\ua\da\da}\rangle$
&& $C$ && $C$ \\
$\langle{\ua\da\da\ua \left|{H_a}\right| \ua\da\da\ua}\rangle$
&& $C$ && $C$ \\
$\langle{\da\da\ua\ua \left|{H_a}\right| \da\da\ua\ua}\rangle$
&& $C$ && $C$ \\
$\langle{\da\ua\ua\da \left|{H_a}\right| \da\ua\ua\da}\rangle$
&& $C$ && $C$ \\
$\langle{\ua\da\ua\da \left|{H_a}\right| \ua\da\ua\da}\rangle$
&& $h/2+C$ && $-h/2+C$ \\
$\langle{\da\ua\da\ua \left|{H_a}\right| \da\ua\da\ua}\rangle$
&& $-h/2+C$ && $h/2+C$
\end{tabular} \end{ruledtabular}
\end{table}

Turning first to the closed set of C and J diagrams, Fig.~\ref{fig:xyvert}, 
we construct the directed-loop equations, which are now different from the 
forms Eq.~(\ref{eqset1}) and Eq.~(\ref{hdirloopA}).
\begin{eqnarray}
a_{11}+a_{12}+a_{13}+a_{14} &=&  C \nonumber, \\ 
a_{21}+a_{22}+a_{23}+a_{24} &=&  \mp h/4+C \nonumber, \\ 
a_{31}+a_{32}+a_{33}+a_{34} &=& {J}/{2} , \label{hsdirloopA} \\ 
a_{41}+a_{42}+a_{43}+a_{44} &=&  {J}/{2} . \nonumber 
\end{eqnarray}
Notice that we have suppressed the explicit definition used before for 
the weights, $W$, in order to simplify notation.  The $\mp$ sign defines 
the convention that the corresponding term is negative if it falls on an 
A plaquette, and positive if it falls on a B plaquette.  We can set the 
bounce processes $a_{14}=a_{24}=0$ as long as $J>h/4$, giving a solution:
\begin{widetext} 
\begin{equation} 
\begin{array}{llll}
a_{11} = C-J/2 \mp h/8, &a_{12} = J/4 \pm h/16, &a_{13} = J/4 \pm h/16,
&a_{14} = 0, \\
a_{21} = C-J/2 \mp h/8, &a_{22} = J/4 \mp h/16, &a_{23} = J/4 \mp h/16,
&a_{24} = 0, \\
a_{31} = J/4 \pm h/16, &a_{32} = J/4 \mp h/16, &a_{33} = 0, &a_{34} = 0, \\
a_{41} = J/4 \pm h/16, &a_{42} = J/4 \mp h/16, &a_{43} = 0, &a_{44} = 0.
\end{array}
\label{ESta1}
\end{equation}
Again, the upper symbol of $\pm$ or $\mp$ refers to the vertex weight on
the A sublattice, and the lower symbol corresponds to the B sublattice.
Note, to keep $a_{11}>0$ on all plaquettes, we require $C \geq J/2+h/8$.

Another solution for the A and B sublattices, valid for $h>4J$, is 
obtained by necessarily requiring some non-zero bounce probabilities.
Consider first the set of equations representing B-plaquettes in
Eq.~(\ref{hsdirloopA}).  A simple solution can be found with a non-zero 
bounce, $a_{24}=h/4-J$.  Note, however,  that this sign in the first term 
in $a_{24}$ that makes this solution invalid on A-plaquettes 
(where $h \rightarrow -h$).  Thus, for $h>4J$ on B plaquettes:
\begin{equation}
\begin{array}{llll}
a_{11} = C, &a_{12} = 0, &a_{13} = 0, &a_{14} = 0,  \\
a_{21} = C, &a_{22} = J/2, &a_{23} = J/2, &a_{24} = h/4-J, \\
a_{31} = 0, &a_{32} = J/2, &a_{33} = 0, &a_{34} = 0, \\
a_{41} = 0, &a_{42} = J/2, &a_{43} = 0, &a_{44} = 0
\end{array}
\label{ESta3}
\end{equation}
For A-plaquettes, another form of the solution is needed, which is 
analogous to the solution  Eq.~(\ref{ESv4}) found for the diagonal 
interaction $V<-J<0$.  Setting $a_{14}=h/2-2J$, and $a_{24}=h/4-J$  
constrains the equations to give the solution ($h>4J$ on A plaquettes):
\begin{equation}
\begin{array}{llll}
a_{11} = C+J-h/2, &a_{12} = J/2, &a_{13} = J/2, &a_{14} = h/2-2J,  \\
a_{21} = C+J-h/2, &a_{22} = 0, &a_{23} = 0, &a_{24} = h/4-J, \\
a_{31} = J/2, &a_{32} = 0, &a_{33} = 0, &a_{34} = 0, \\
a_{41} = J/2, &a_{42} = 0, &a_{43} = 0, &a_{44} = 0.
\end{array}
\label{ESta4}
\end{equation}

\begin{figure}
\includegraphics[height=5.5cm]{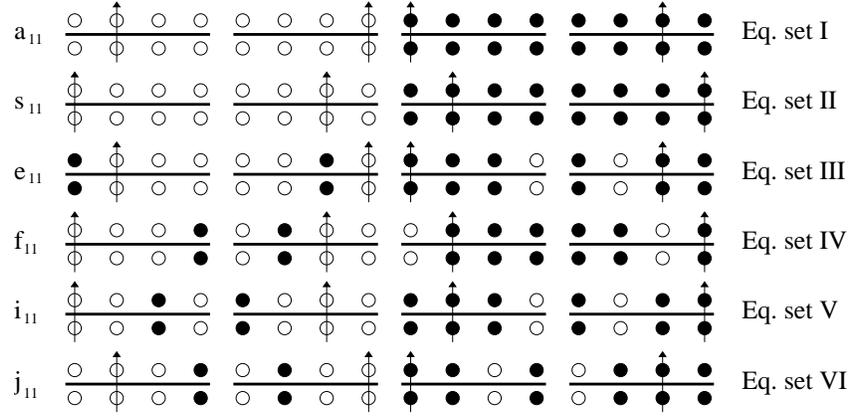}
\caption{Reference elements of the closed C and J vertex-path 
diagrams.  The element letters (left) and equation set designation (right)
refer to the relevant vertex weight equation set, Tables~\ref{EqI}-\ref{EqVI}.}
\label{redST}
\end{figure}

\end{widetext}
This imposes the requirement that $C>h/2-J$.

\begin{center}
\begin{figure}
\includegraphics[height=4cm]{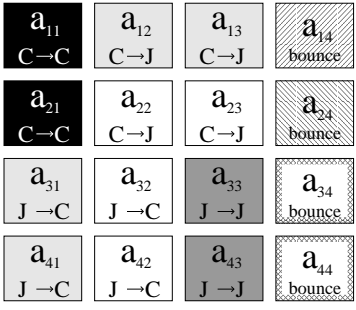}
\caption{Schematic representation of the closed set of C and J 
vertex-paths used in solving the directed-loop equations for the  J-K 
model.  Blocks with the same shading represent equivalent vertex weights. }
\label{Ssymm}
\end{figure}
\end{center}

This outlines the basic method of constructing the directed-loop probabilities 
for the staggered magnetic field Hamiltonian.  The only difficulty in 
completing the procedure is identifying all of the separate sets of closed 
vertex-path diagrams which contribute different vertex weights to the directed 
loop algorithm.  Instead of explicitly illustrating and solving all of these 
different sets in this case, we simply present the solutions in a more concise 
form.  To begin, note that we can abbreviate the illustration of the closed 
sets of C and J vertex-paths if we constrain the solutions to obey the 
same detailed balance, Eq.~(\ref{balancereq}), and symmetry arguments, 
Eq.~(\ref{symmv}), as used throughout this paper.  In this case, one only 
needs to know the upper-left (reference) vertex ($a_{11}$ in 
Fig.~\ref{fig:xyvert}) 
in order to uniquely define the entire closed set of C to J vertex paths.  
The rules for constructing the closed set, as discussed in section II E, 
can then be summarized  by the schematic representation in Fig.~\ref{Ssymm}, 
which illustrates the general relationship between the different vertex 
weights, and their corresponding transformations.
We can therefore easily construct an entire closed set of C and J vertex 
paths using Fig.~\ref{Ssymm} simply by defining the reference vertex. 
Following this procedure, we see that the number of unique vertex probability
solution sets is narrowed down to six, unrelated by trivial symmetry operations.
The reference vertices for these six unique sets are illustrated in 
Fig.~\ref{redST}.  The corresponding vertex weights are summarized in the 
equation sets of Tables \ref{EqI}-\ref{EqVI}.  For example, the first 
reference vertex of Eq. set I in Fig.~\ref{redST} corresponds to $a_{11}$ of
Fig.~(\ref{fig:xyvert}).  The corresponding vertex weights appear in 
Table \ref{EqI}, and are equivalent to the solution sets 
Eqs.~(\ref{ESta1}), (\ref{ESta3}) and (\ref{ESta4}).  The tables are 
abbreviated to only include unique weights not related by the symmetries 
of Fig.~\ref{Ssymm}, however the full equation sets are recovered easily 
by using this figure or Eqs.~(\ref{balancereq}) and  (\ref{symmv}).

\begin{center}
\begin{table}
\caption{Vertex weight equation, set I, for the staggered-field J-K model.  
An example of the starting vertex, $a_{11}$, is illustrated in 
Fig.~\ref{redST}.}
\label{EqI}
\begin{ruledtabular}
\begin{tabular}{cp{4mm}cp{4mm}cp{4mm}c}
vertex&&A/B ($h<4J$) &&A ($h>4J$)&&B ($h>4J$)\\ \hline
$a_{11}$ && $C-J/2 \mp h/8$  && $C+J-h/2$ && $C$ \\
$a_{12}$ && $J/4 \pm h/16$ && $J/2$ && $0$ \\
$a_{14}$ && 0  && $h/2-2J$ && $0$\\
$a_{22}$ && $J/4 \mp h/16$  &&  0 && $J/2$\\
$a_{24}$ && 0 && $h/4-J$ && $h/4-J$ \\
$a_{33}$ && 0 && $0$ && 0 \\
$a_{34}$ && 0  && 0 && 0
\end{tabular} \end{ruledtabular}
\end{table}

\begin{table}
\caption{Vertex weight equation set II.}
\label{EqII}
\begin{ruledtabular}
\begin{tabular}{cp{4mm}cp{4mm}cp{4mm}c}
vertex&&A/B ($h<4J$) &&A ($h>4J$)&&B ($h>4J$)\\ \hline
$s_{11}$ && $C-J/2 \pm h/8$  && $C$ && $C+J-h/2$ \\
$s_{12}$ && $J/4 \mp h/16$  && $0$ && $J/2$ \\
$s_{14}$ && 0  && 0 && $h/2-2J$ \\
$s_{22}$ && $J/4 \pm h/16$  && $J/2$ && 0 \\
$s_{24}$ && 0  && $h/4-J$ && $h/4-J$ \\
$s_{33}$ && 0  && 0 && $0$ \\
$s_{34}$ && 0  && 0 && 0
\end{tabular} \end{ruledtabular}
\end{table}

\begin{table}
\caption{Vertex weight equation set III.}
\label{EqIII}
\begin{ruledtabular}
\begin{tabular}{cp{4mm}cp{4mm}cp{4mm}c}
vertex&&A/B ($h<4J$) &&A ($h>4J$)&&B ($h>4J$)\\ \hline
$e_{11}$ && $C-J/2 \pm h/8$ && $C$ && $C+J-h/2$ \\
$e_{12}$ && $J/4 \pm h/16$  && $J/2$ && $0$ \\
$e_{14}$ && 0  && $h/4-J$ && $h/4-J$ \\
$e_{22}$ && $J/4 \mp h/16$ && $0$ && $J/2$ \\
$e_{24}$ && 0 && 0 && $h/2-2J$ \\
$e_{33}$ && 0 && 0 && $0$ \\
$e_{34}$ && 0 && 0 && 0
\end{tabular} \end{ruledtabular}
\end{table}

\begin{table}
\caption{Vertex weight equation set IV.}
\label{EqIV}
\begin{ruledtabular}
\begin{tabular}{cp{4mm}cp{4mm}cp{4mm}c}
vertex&&A/B ($h<4J$) &&A ($h>4J$)&&B ($h>4J$)\\ \hline
$f_{11}$ && $C-J/2 \mp h/8$  && $C+J-h/2$ && $C$ \\
$f_{12}$ && $J/4 \mp h/16$  && $0$ && $J/2$ \\
$f_{14}$ && 0 && $h/4-J$ && $h/4-J$ \\
$f_{22}$ && $J/4 \pm h/16$  && $J/2$ && 0 \\
$f_{24}$ && 0  && $h/2-2J$ && 0 \\
$f_{33}$ && 0 && $0$ && $0$ \\
$f_{34}$ && 0  && 0 && 0
\end{tabular} \end{ruledtabular}
\end{table}

\begin{table}
\caption{Vertex weight equation set V.}
\label{EqV}
\begin{ruledtabular}
\begin{tabular}{cp{4mm}cp{4mm}cp{4mm}c}
vertex&&A/B ($h<4J$) &&A ($h>4J$)&&B ($h>4J$)\\ \hline
$i_{11}$ && $C-J/2 \pm 3h/8$ && $C+h/4$ && $C+J-3h/4$ \\
$i_{12}$ && $J/4 \mp h/16$ && $0$ && $J/2$ \\
$i_{14}$ && 0  && $0$ && $h/2-2J$ \\
$i_{22}$ && $J/4 \pm h/16$  && $J/2$ && 0 \\
$i_{24}$ && 0 && $h/4-J$ && $h/4-J$ \\
$i_{33}$ && 0 && 0 && $0$ \\
$i_{34}$ && 0 && 0 && 0
\end{tabular} \end{ruledtabular}
\end{table}

\begin{table}
\caption{Vertex weight equation set VI.}
\label{EqVI}
\begin{ruledtabular}
\begin{tabular}{cp{4mm}cp{4mm}cp{4mm}c}
vertex&&A/B ($h<4J$) &&A ($h>4J$)&&B ($h>4J$)\\ \hline
$j_{11}$ && $C-J/2 \mp 3h/8$ && $C+J-3h/4$ && $C+h/4$ \\
$j_{12}$ && $J/4 \pm h/16$ && $J/2$ && $0$ \\
$j_{14}$ && 0 && $h/2-2J$ && $0$\\
$j_{22}$ && $J/4 \mp h/16$ &&  0 && $J/2$\\
$j_{24}$ && 0  && $h/4-J$ && $h/4-J$ \\
$j_{33}$ && 0 && $0$ && 0 \\
$j_{34}$ && 0 && 0 && 0
\end{tabular} \end{ruledtabular}
\end{table}
\end{center}

\section{Discussion}

In summary, we have developed in this work an extensive algorithmic framework 
for SSE quantum Monte Carlo simulations of the $S=1/2$ XY model with ring 
exchange -- the J-K model -- on a 2D square lattice.
In addition to outlining the basic representation of the quantum mechanical
partition function as a power series expansion of plaquette-operators 
acting on a chosen basis in the $S^z$ representation, we have developed
advanced implementations of the directed-loop and multi-branch cluster updates,
designed to significantly increase algorithm efficiency in various parameter
regimes of the Hamiltonian.
We have studied the performance of the various updating procedures using
autocorrelation functions. 
We have also outlined
modifications of the directed-loop equations to account for extensions
of the J-K Hamiltonian to include diagonal (potential energy) operators.  
Although several specific Hamiltonian terms are discussed, the
procedure developed is sufficiently general to allow for easy extensions to
other diagonal interactions.

\begin{table}[ht]
\caption{\label{MCexactD}
Comparison of ground-state energy (per spin) of exact diagonalization and
SSE quantum Monte Carlo results for various parameter values of the
J-K Hamiltonian, on a $4 \times 4$ square lattice.
The exchange was set to $J=1/2$, and simulations were
performed with 50 million Monte Carlo production steps each.
The staggered field strength is represented by $h_s$.
}
\begin{ruledtabular}
\begin{tabular}{llllll}
$K/J$  & $V/J$ & $h/J$   & $h_s/J$ & $E_{exact}$ & $E_{qmc}$   \\ \hline
0 & 0 & 0 & 0 & -0.5624863 & -0.56249(1)    \\
1 & 0 & 0 & 0 & -0.6803518 & -0.68034(1)   \\
4 & 0 & 0 & 0 & -1.1530991 & -1.15311(2) \\
0 & $1/2$ & 0 & 0 & -0.6239222 & -0.62392(1)   \\
2 & 2 & 0 & 0 & -1.2864452 & -1.28643(2) \\
1 & -3 & 0 & 0 & -0.3983951 & -0.39838(1) \\
1 & 0 & 2 & 0 & -0.7434355 & -0.74343(1)\\
5 & 0 & 6 & 0 & -1.5000000 & -1.50001(1) \\
4 & 0 & 0 & 2 & -1.2547499 & -1.25476(2) \\
1 & 0 & 0 & 5 & -1.3859171 & -1.38590(1) \\
\end{tabular}
\end{ruledtabular}
\end{table}

The last step needed to provide confidence in the rather complex implementation 
of our directed-loop algorithms discussed here is to carry out rigorous testing. 
We do this by comparing SSE data with results obtained by exact diagonalization 
of the Hamiltonian. Table~\ref{MCexactD} compares 
ground-state energies obtained in various algorithmic solution regimes of the 
quantum Monte Carlo schemes discussed here with exact diagonalization results
for $4\times 4$ lattices. In the simulations, the temperature $T_0$ was chosen 
sufficiently low for there to be no differences, within statistical errors, 
between simulations carried out at $T_0$ and $2T_0$. The absence of any detectable 
differences between the exact and SSE results to a relative statistical
accuracy of $\approx 10^{-5}$ illustrates the unbiased nature of these 
calculations.

The SSE algorithm developed here can be extended straightforwardly to J-K 
models with four-spin exchange terms on other lattices. For example, implementation 
of the Hamiltonian on the triangular\cite{LAtri} and kagome\cite{balents} 
lattices is possible for some parameter regimes without being hampered by
the sign problem.  In particular, quantum Monte Carlo simulations at
or near the RK-point\cite{RK} ($J=0$, $-K=V$) are anticipated to make explicit 
connection with predictions from analytical theories.  Studies on these models 
are underway, and are expected to reveal a rich variety of ground state phenomenon.

In principle, the scheme can be extended also to
multi-spin interactions on rings with more than four spins, 
for example the XY model with six-spin exchange on the pyrochlore 
lattice.\cite{Mike1}
However, it is clear that the directed-loop scheme will then become quite 
complex, and explicit solutions of the type we have presented here may not
be practical.

%\null\vskip1mm

\acknowledgments{
We would like to thank Leon Balents, Matthew Fisher, and Doug Scalapino 
for valuable discussions. We are grateful to Luca Capriotti for providing some 
exact diagonalization codes. RGM acknowledges support from the Department of Energy,
Grant No.~DOE85-45197. AWS would like to thank the Department of Physics at
UCSB for hospitality during a visit. Part of the work was also carried out at the 
Kavli Institute for Theoretical Physics, supported under NSF Grant No.~PHY99-07949.}

\null

\null\vskip5cm

\end{document}